\documentclass[aps,prx,10pt, twocolumn, superscriptaddress, amsmath, amssymb,longbibliography]{revtex4-2}
\usepackage[british]{babel}

\usepackage{graphicx}   
\usepackage{bm}          
\usepackage{verbatim}  
\usepackage{times}

\usepackage{mathtools}
\usepackage{physics} 
\usepackage{siunitx}
\usepackage[version=4]{mhchem}
\usepackage[dvipsnames]{xcolor}
\definecolor{mygreen}{RGB}{44,85,17}
\definecolor{myblue}{RGB}{34,31,150}
\definecolor{myred}{RGB}{255,66,56}

\usepackage[breaklinks=true,colorlinks=true,linkcolor=myblue,urlcolor=myblue,citecolor=myblue]{hyperref}

\begin{document}



\title{Optimal cold atom thermometry using adaptive Bayesian strategies}

\author{Jonas Glatthard}
\email{J.Glatthard@exeter.ac.uk}
\affiliation{Department of Physics and Astronomy, University of Exeter, Stocker Road, Exeter EX4 4QL, United Kingdom.}

\author{Jes\'{u}s Rubio}
\email{J.Rubio-Jimenez@exeter.ac.uk}
\affiliation{Department of Physics and Astronomy, University of Exeter, Stocker Road, Exeter EX4 4QL, United Kingdom.}

\author{Rahul Sawant}
\affiliation{School of Physics and Astronomy, University of Birmingham, Edgbaston, Birmingham B15 2TT, United Kingdom.}

\author{Thomas Hewitt}
\affiliation{School of Physics and Astronomy, University of Birmingham, Edgbaston, Birmingham B15 2TT, United Kingdom.}

\author{Giovanni Barontini}
\email{G.Barontini@bham.ac.uk}
\affiliation{School of Physics and Astronomy, University of Birmingham, Edgbaston, Birmingham B15 2TT, United Kingdom.}

\author{Luis A. Correa}
\affiliation{Department of Physics and Astronomy, University of Exeter, Stocker Road, Exeter EX4 4QL, United Kingdom.}

\date{\today}
   

\begin{abstract}

Precise temperature measurements on systems of few ultracold atoms is of paramount importance in quantum technologies, but can be very resource-intensive. Here, we put forward an adaptive Bayesian framework that substantially boosts the performance of cold atom temperature estimation. Specifically, we process data from real and simulated release--recapture thermometry experiments on few potassium atoms cooled down to the microkelvin range in an optical tweezer. From simulations, we demonstrate that adaptively choosing the release--recapture times to maximise information gain does substantially reduce the number of measurements needed for the estimate to converge to a final reading. Unlike conventional methods, our proposal systematically avoids capturing and processing uninformative data. We also find that a simpler non-adaptive method exploiting all the \textit{a priori} information can yield competitive results, and we put it to the test on real experimental data. Furthermore, we are able to produce much more reliable estimates, especially when the measured data are scarce and noisy, and they converge faster to the real temperature in the asymptotic limit. Importantly, the underlying Bayesian framework is not platform-specific and can be adapted to enhance precision in other setups, thus opening new avenues in quantum thermometry.

\end{abstract}

\maketitle


\section{Introduction}

Precise low-temperature thermometry is a key enabler for cold atom-based quantum technologies. In particular, analog quantum simulation in optical lattices \cite{sherson2010single,bloch2012quantum}, computation on large-scale programmable simulators \cite{ebadi2021quantum,scholl2021quantum}, the study of thermalisation in closed quantum systems \cite{langen2013local,langen2015ultracold,bouton2020single}, or the realisation of thermodynamic cycles \cite{niedenzu2019quantized,barontini2019ultra,bouton2021quantum}. The most common thermometric technique for cold atoms is time-of-flight imaging, which infers temperature from the velocity distribution of a free-expanding atomic cloud. 
However, this is not sufficiently accurate in optical lattices. It also becomes unsuitable in more advanced setups like optical tweezers, which allow us to independently trap and manipulate only one or few atoms \cite{serwane2011deterministic,kaufman2012cooling,li2012crossed,ebadi2021quantum,scholl2021quantum}. In these cases, alternatives such as Raman sideband spectroscopy \cite{kaufman2012cooling}, release--recapture thermometry \cite{chu1985three,lett1988observation,mudrich2002cooling}, or selective atom spilling \cite{serwane2011deterministic} must be employed. Often, however, such techniques require a large number of experimental repetitions to be accurate, rendering thermometry expensive in terms of machine time.

On the theoretical side, the application of estimation-theoretic methods to low-temperature thermometry has consolidated into the novel field of `quantum thermometry' \cite{DePasquale2018,mehboudi2019review}. Specifically, progress has been made on establishing fundamental scaling laws for the signal-to-noise ratio of low-temperatur estimates \cite{hovhannisyan2018,potts2019fundamental,henao2021thermometric}, or on the identification of design prescriptions that can make a probe more responsive to temperature fluctuations \cite{correa2015probe,plodzie2018fermion,mukherjee2019thermometry_control,mitchison2020situ,glatthard2022bending,correa2017thermometry_strongcoupling,seah2019collisional}. As a result, precision tuning in sensing applications with atomic impurities is starting to be informed by estimation theory \cite{bouton2020single,nettersheim2022sensitivity,mehboudi2019using}.

Most of the existing literature on quantum thermometry relies on \textit{local} estimation theory \cite{mehboudi2019review,DePasquale2018}, and adopts the (quantum) Fisher information $ \mathcal{F} $ as the figure of merit to be maximised \cite{braunstein1994statistical}. Indeed, in the asymptotic limit of large number of measurements (i.e., $ \mu \rightarrow \infty $), and for unbiased estimators, $ \mathcal{F} $ controls the scaling of the best-case signal-to-noise ratio of temperature estimates, as per $(T/\delta T)^2 \sim \mu\,\mathcal{F}\,T^2$. However, in the experimentally relevant scenario of \textit{finite} measurement records, with just tens or hundreds of shots, $ \mathcal{F} $ does not always capture---even qualitatively---the behaviour of optimal temperature estimates. In these cases, one must adopt the more general Bayesian framework \cite{helstrom1976book, jaynes2003, toussaint2011}. This has recently led to a wealth of new results in quantum thermometry \cite{rubio2021global,mok2021optimal,mehboudi2021fundamental,jorgensen2021bayesian,boeyens2021noninformative,rubio2021scales,alves2021bayesian}, although there had been earlier applications \cite{prosper1993,johnson2016thermometry}. 

While experimental proposals haven been put forward within quantum thermometry \cite{bouton2020single,plodzie2018fermion,mitchison2020situ,mehboudi2019using,higgins2014superabsorption,mehboudi2015thermometry}, the field has remained eminently theoretical and mostly concerned with setting ultimate precision bounds \cite{mehboudi2019review}. With this work we want to take a more practical turn by exploiting the theory to process actual experimental data. This has been facilitated by the adoption of the Bayesian framework which, not only provides precision bounds as in local estimation theory, but also allows for the construction of optimal estimators from measurement outcomes \cite{helstrom1976book, jaynes2003, toussaint2011, rubio2021scales}.

By exploiting Bayesian methods, here we develop an adaptive strategy to optimally assess the temperature of single- or few-atom systems. Namely, given a set of measurement outcomes and a model for the experiment, we find the estimate which minimises, on average, a suitably defined error. Crucially, we can compute the expected information gain shot by shot, which allows us to adjust the experiment in real time so that each measurement is maximally informative. 

We illustrate the benefits of our adaptive Bayesian protocol by optimising a release--recapture thermometry experiment on a few \ce{^{41}K} atoms tightly confined in an optical tweezer at \SI{}{\micro\kelvin} temperatures. This is a particularly challenging system in which precise thermometry is usually exceedingly expensive in terms of the number of measurements needed. Indeed, as already advanced, single-shot procedures like time-of-flight absorption imaging cannot be used with a single or few atoms, due to the vanishing optical density. On the other hand, alternative methods like Raman sideband spectroscopy operate at far lower temperatures, which  leaves release and recapture as the only option. Our aim here is to show how resource optimisation can help to overcome the two main weaknesses of release--recapture thermometry; namely, that it inherently needs many experimental repetitions, and that it is usually less accurate than other methods. To do so, we will use the light-off time of free expansion of the atoms as the control parameter in our optimisation. 

Working with both real and simulated data, we show that our adaptive Bayesian protocols largely outperform conventional release--recapture thermometry in practice. Specifically, our methods showcase a much reduced estimate variability when applied to finite measurement records with equal number of data points. Our approach is thus more dependable. Furthermore, we find that, in the asymptotic limit of many data, the estimates from our Bayesian protocols converge faster to the true temperature than conventional estimates. Finally, we show how guiding data acquisition by information gain can substantially reduce the number of measurements needed for an estimate to converge to a stable value, thus helping to optimise resources. Ultimately, in this work we demonstrate in practice how quantum thermometry can provide concrete design prescriptions capable of making a difference in experiments. 

This paper is structured as follows: In Sec.~\ref{sec:bayes-paradigm} we introduce our Bayesian method. In Sec.~\ref{sec:release--recapture} we describe a simple analytical model for the recapture probability. Next, in Sec.~\ref{sec:maximising-info}, we describe three different thermometric protocols and compare their performance. These consist in the mere Bayesian processing of data captured at unoptimised times, the repeated measurement at the \textit{a priori} optimal recapture time, and a fully adaptive method. Details on our experimental setup and on how real data are processed are given in Sec.~\ref{sec:real-experimental}. In Sec.~\ref{sec:results}, we rank the protocols in terms of estimate variability and asymptotic convergence speed and show how information-maximisation reduces the required number of measurements for estimates to converge to a target precision. Finally, in Sec.~\ref{sec:conclusion} we draw our conclusions.

\section{Adaptive Bayesian thermometry}
\label{sec:bayesian}

\subsection{Bayesian data analysis and global thermometry}
\label{sec:bayes-paradigm}

We start by outlining the Bayesian--global paradigm introduced by some of us in Refs.~\cite{rubio2021global,rubio2021scales}. Even though we later focus on the specific release--recapture setup, this framework is fully applicable to any thermometry experiment in which temperature behaves as a `scale parameter' \cite{jaynes1968, rubio2021scales}. 
For a more detailed account of this paradigm and why it applies to release--recapture thermometry, see Appendix~\ref{sec:scale-est-theory}.

Our aim is to estimate a temperature $ T $ from a set of measurement outcomes $\boldsymbol{n} = (n_1, \dots, n_\mu)$ recorded in $\mu$ runs of a given experiment. In addition, we consider a controllable parameter $ t $ that we set to $\boldsymbol{t} = (t_1, \dots, t_\mu)$ in each subsequent run.  
In order to extract maximum information from these data, we need a likelihood function \cite{jaynes2003, toussaint2011}.
Assuming all shots to be independent, this amounts to finding a probability $p(n_i|T,t_i)$ for obtaining the outcome $ n_i $ given $t_i$ and $T$. 
Furthermore, since the true temperature is unknown, we work with a hypothesis $\theta \in [\theta_\mathrm{min}, \theta_\mathrm{max}]$ about its value.

The idea is then to assign a weight to the different $\theta$ in light of the measurement record $(\boldsymbol{n},\boldsymbol{t})$. This is achieved by constructing the probability $p(\theta|\boldsymbol{n}, \boldsymbol{t})$ \cite{jaynes2003,toussaint2011}. 
Using Bayes theorem, such distribution can be cast as (cf. Appendix~\ref{app:posterior-derivation})
\begin{equation}
    p(\theta|\boldsymbol{n},\boldsymbol{t}) \propto p(\theta) \prod\nolimits_{i=1}^\mu p(n_i|\theta, t_i),
    \label{eq:posterior-all-data}
\end{equation}
where $p(n_i|\theta,t_i)$ is our likelihood model evaluated at $\theta$, and $p(\theta)$ is a probability representing information about $T$ \textit{prior} to collecting the data. 

Since one does not, in general, have any prior knowledge about $T$, we choose the distribution $p(\theta)$ corresponding to maximum ignorance within $[\theta_\mathrm{min}, \theta_\mathrm{max}]$ \cite{jaynes2003, demkowicz2020}. 
As we shall see next, the energy scale in which our model $ p(n_i\vert T,t_i) $ for release--recapture thermometry is applicable is determined by the unknown temperature itself. We thus say that temperature is a scale parameter \cite{jaynes1968,prosper1993,rubio2021global, rubio2021scales}, and adopt Jeffreys's prior
\begin{equation}\label{eq:prior}
    p(\theta) = \left[\theta\,\log{\left(\frac{\theta_{\mathrm{max}}}{\theta_{\mathrm{min}}}\right)}\right]^{-1}
\end{equation}
to represent maximum ignorance about it \cite{jeffreys1961,jaynes2003,toussaint2011}.

Having fixed $p(\theta)$ and $p(n_i,t_i|\theta)$, Eq.~\eqref{eq:posterior-all-data} is used as follows. 
Upon recording $(n_1,t_1)$, the likelihood $p(n_1|\theta,t_1)$ is calculated as a function of $\theta$. 
The functions $p(\theta)$ and $p(n_1|\theta,t_1)$ are multiplied and the result normalised---this operation is interpreted as having \emph{updated} the initial information with the first pair of data. 
Next, $(n_2,t_2)$ is recorded and the resulting probability, multiplied by $p(n_2|\theta,t_2)$ and normalised. 
After $\mu$ iterations, one finally arrives at $p(\theta|\boldsymbol{n},\boldsymbol{t})$, which does encode all the available temperature information. That is why $p(\theta|\boldsymbol{n},\boldsymbol{t})$ is referred-to as `posterior probability'.

At this point, one can already extract an estimate and its associated error. 
As shown in Ref.~\cite{rubio2021global}, whenever $T$ behaves as a scale parameter, the \textit{optimal} estimator is given by
\begin{equation}
    \tilde{\vartheta}(\boldsymbol{n},\boldsymbol{t}) = \theta_u \exp\left[\int d\theta \,p(\theta|\boldsymbol{n},\boldsymbol{t}) \log{\left(\frac{\theta}{\theta_u}\right)}\right],
    \label{eq:est}
\end{equation}
where $\theta_u$ simply neutralises the units within the logarithm \cite{matta2011} without affecting the value of $\tilde{\vartheta}(\boldsymbol{n},\boldsymbol{t})$. 
In this work, we set $\theta_u = \SI{1}{\micro\kelvin}$.
In turn, the `error bar' on the estimator \eqref{eq:est} is 
\begin{subequations}\label{eq:error-bar-and-mle}
\begin{equation}
    \Delta \tilde{\vartheta}(\boldsymbol{n},\boldsymbol{t}) =  \tilde{\vartheta}(\boldsymbol{n},\boldsymbol{t}) \sqrt{\bar{\epsilon}_{\mathrm{mle}}(\boldsymbol{n},\boldsymbol{t})},
    \label{eq:error-bar}
\end{equation}
where
\begin{equation}
    \bar{\epsilon}_{\mathrm{mle}}(\boldsymbol{n},\boldsymbol{t}) = \int d\theta\,p(\theta|\boldsymbol{n},\boldsymbol{t}) \log^2{\left[\frac{\tilde{\vartheta}(\boldsymbol{n},\boldsymbol{t})}{\theta}\right]}
    \label{eq:mle-experiments}
\end{equation}
\end{subequations}
is the mean logarithmic error---a relative error---evaluated for the specific dataset at hand. Ultimately, we report Bayesian temperature estimates as $\tilde{\vartheta}(\boldsymbol{n}, \boldsymbol{t}) \pm \Delta \tilde{\vartheta}(\boldsymbol{n}, \boldsymbol{t})$.

We will also exploit the notion of mean information gain for a single shot. 
As per Ref.~\cite{rubio2021global}, this can be quantified as
\begin{equation}
    \mathcal{K}(t)=\sum\nolimits_{n} p(n|t) \log^2{\left[\frac{\tilde{\vartheta}(n,t)}{\tilde{\vartheta}_p}\right]},
    \label{eq:bayes-info}
\end{equation}
where 
\begin{subequations} \label{eq:bayes-info-pieces}
\begin{align}
&\tilde{\vartheta}_p = \theta_u \exp\left[\int d\theta \,p(\theta) \log{\left(\frac{\theta}{\theta_u}\right)}\right], \\
&p(n|t) = \int d\theta\,p(\theta)\,p(n|\theta, t).
\end{align}
\end{subequations}
This will allow us to adaptively adjust the control parameter $ t $ so as to maximise $ \mathcal{K}(t) $ at every step, thus making each measurement maximally informative.

The Bayesian--global paradigm summarised here will suffice to optimise release--recapture experiments. Nevertheless, we note that alternative formulations of Bayesian thermometry \cite{jorgensen2021bayesian, boeyens2021noninformative, pearce2017,mok2021optimal,alves2021bayesian} exist. 
For a perspective on when it is appropriate to employ these, see Refs.~\cite{rubio2021global, rubio2021scales}.

\subsection{The release--recapture method}
\label{sec:release--recapture}

In a release-recapture thermometry experiment, the trap confining an ultracold gas at temperature $T$ is switched off during a time $t$, so that the atoms expand ballistically. 
Upon reactivating the trap, the number of recaptured atoms $n$ is measured by fluorescence imaging. The temperature is obtained by fitting measurement outcomes at different times to a model $f(T,t)$ for the fraction of recaptured atoms \cite{mudrich2002cooling}. 

Let us start with the simple case of a trap deterministically loaded with one atom, deferring the multi-atom scenario to Sec.~\ref{sec:model}. In that case, the probabilities for successfully recapturing or losing the atom after time $t_i$ are, respectively 
\begin{subequations}\label{eq:model}
\begin{align}
p(n_i = 1|T,t_i) &= f(T,t_i), \label{eq:model-survival} \\    
p(n_i = 0|T,t_i) &= 1 - f(T,t_i);
\end{align}
\end{subequations}
this gives us the required likelihood model.

Next, we need to calculate the recaptured fraction $f(T,t)$. 
The trapping potential is created by focusing a laser beam to a narrow waist, which produces tight confinement in the transverse $x$--$y$ plane. This is described by 
\begin{equation}
    U(r) = -U_0\,\mathrm{exp}\left(-\frac{2 r^2}{w_0^2}\right),
    \label{eq:trap-pot}
\end{equation}
where $U_0$ is the trap depth, $w_0$ is the beam waist, and $r^2 = x^2 + y^2$.
At low $T$, i.e., $U_0/k_B T \gg 1$, Eq.~\eqref{eq:trap-pot} leads to \cite{mudrich2002cooling, mudrich203phd} (cf. Appendix~\ref{app:survival})
\begin{equation}
    f(T,t) = \frac{1}{g(\eta)}\,g{\left[\frac{\eta\,W(\tilde{t}^2)}{\tilde{t}^2}\right]},
    \label{eq:recapture}
\end{equation}
where $g(s) = 1 - \mathrm{e}^{-s}$, $\eta=U_0/(k_B T)$, $\tilde{t}^2=4 U_0 t^2/(m w_0^2)$, $m$ is the mass of the atom, and $W(\cdot)$ is the Lambert function, defined implicitly as  $W(s)\,\mathrm{e}^{W(s)}\coloneqq s$. 
By resorting to numerical simulations, one could further account for the axial motion of the atoms, i.e., their initial position distribution and the effect of gravity. However, Eq.~\eqref{eq:recapture} is in very good agreement with such simulations in the low-$T$ regime (see Appendix~\ref{app:survival}). 

To estimate $ T $ in practice, one conventionally chooses a range of free-expansion times $\lbrace t_1,\cdots,t_\nu\rbrace$ so that the decay of $f(T,t)$ is well sampled \cite{chu1985three,lett1988observation}. At each time $t_i$, $\alpha_i$ independent photon-count measurements are performed, so as to infer the number of recaptured atoms $ \{n_{ij}\}_{j=1}^{\alpha_i} $. Eventually, one records the average $ \overline{n}_i $ for every $ t_i $, and fits these to $\lambda f(T,t)$ via least squares. The light-off time $t$ then acts as independent variable, while $T$ and $\lambda$ are fitting parameters \cite{mudrich2002cooling}.

In spite of the simplicity of the method, two issues must be noted. First, assuming that the empirical average $\bar{n}_i$ is close to its true value presupposes an indeterminate `large' number of measurements $\alpha_i$. Hence, should $ \alpha_i $ not be `large enough', the final estimate could become unreliable \cite{braunstein1992,rafal2015}.
Secondly, the method discards most of the information available---the full measurement record $ \{n_{ij}\} $ is known and yet, one chooses to work only with the averages $\{\overline{n}_i\}$, which likely leads to precision loss \cite{jaynes2003}. Here, $ i\in\{ 1,\cdots,\nu \} $ and $ j \in \{ 1,\cdots,\alpha_i \} $. 
Both issues are naturally bypassed by the Bayesian estimation framework of global thermometry \cite{rubio2021global} as discussed below.

\begin{figure*}[t]
\includegraphics[width=0.8\textwidth]{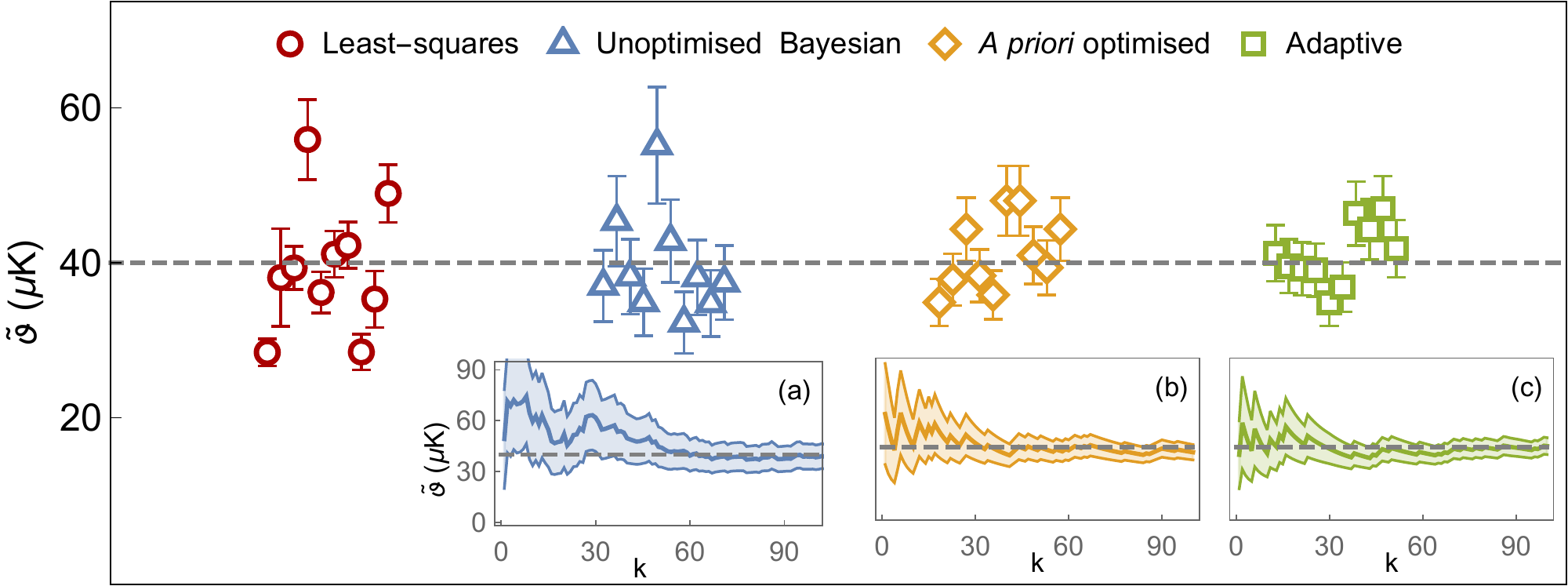}
\caption{\textbf{Reliability of estimation strategies for a trap loaded with a single atom.} 
Temperature estimates and their error bars, obtained from processing simulated release--recapture measurements on a single \ce{^{41}K} atom with different protocols (see Sec.~\ref{sec:singl-atom} for details). The results obtained from $10$ different realisations are shown. The underlying true temperature was set to \SI{40}{\micro\kelvin} (dashed grey). First, a record with $30$ measurements at $7$ unoptimised expansion times was fitted to the recapture probability, as discussed in Sec.~\ref{sec:release--recapture} (red circles), as well as processed with our Bayesian--global technique (blue triangles). The time-optimised methods were also applied to the same overall number of data points (i.e., $210$), simulated, respectively, at the optimal \textit{a priori} recapture time $ t_s $ (orange rhombs), and at adaptively chosen variable optimal times (green squares). Optimisation of the recapture times noticeably reduces the variability of the estimates. Note that there is no abscissa axis in the main plot; the estimates have been squeezed together to help visualise their variability. The insets (a--c) show the typical convergence of global estimates with the number of measurements $k$. As shown, time-optimised methods typically require much less data to converge, when compared with their unoptimised counterparts. 
Also, the final error bars are smaller in (b) and (c).}
\label{fig1}
\centering
\end{figure*}

\subsection{Adaptive maximisation of the information gain}
\label{sec:maximising-info}

We now move on to discuss how Bayesian methods can be employed to process the data form conventional release--recapture experiments and, more importantly, how Bayesian data analysis can inform the measurement strategy. 
Before we proceed, some remarks are in order. 

First, we have just seen that the validity of Eq.~\eqref{eq:recapture} rests on the condition $U_0/(k_B T) \gg 1$, so that it is the unknown $ T $ itself which determines the energy scale in which our model for the fraction of recaptured atoms is applicable.
Hence, we are dealing with a scale-estimation problem \cite{jaynes1968, rubio2021scales} for which Eqs.~(\ref{eq:prior}--\ref{eq:bayes-info}) are appropriate.

As for the range $[\theta_\text{min},\theta_\text{max}]$, a conservative choice for the lower bound is $ 5\%$ of the trap depth in units of the Boltzmann constant (i.e., $U_0/k_B$), as the final cooling stage that the atoms undergo in the trap is not expected to reach such low temperatures (see Sec.~\ref{sec:setup}). On the other hand, we choose $\theta_\mathrm{max}$ as the temperature at which Eq.~\eqref{eq:recapture} starts to break down (see Appendix~\ref{app:survival}).

\subsubsection{Unoptimised Bayesian protocol}\label{sec:unoptimised-bayesian}

Let us consider a release--recapture experiment such as those described in Sec.~\ref{sec:release--recapture}. In this case, the vector $ \boldsymbol{t} $ of expansion times would be
\begin{equation*}
    \boldsymbol{t} = (\smash[t]{\underbrace{t_1, \dots, t_1}_\text{$\alpha_1$ times}},\dots,\smash[t]{\underbrace{t_\nu,\dots,t_\nu}_\text{$\alpha_\nu$ times}}),
\end{equation*}
where the values $ t_i $ would be chosen `by eye' such that the decay of the survival probability \eqref{eq:model} is well sampled. Processing the resulting measurement record $(\boldsymbol{n},\boldsymbol{t})$ with Eqs.~(\ref{eq:prior}--\ref{eq:bayes-info}) may return slightly better estimates than the conventional least-squares fit of the recaptured-atom-averages $ \{\overline{n}_i\} $, since we make use of \textit{all} the available data (see Secs.~\ref{sec:singl-atom} and \ref{sec:res-a}). 

Not all expansion times $ t_i $ are equally informative, however. Intuitively, if the temperature is low, the recapture probability at short enough times is roughly constant, as the atoms have no time to move far---it is those measurements at longer times which carry most of the temperature information. The opposite is true if $ T $ is large. Hence, given that the temperature is unknown, by evenly sampling a time interval we are likely to generate and process many uninformative data points. Instead, as the temperature information is updated by feeding in measurement outcomes, efforts should be concentrated in sampling only those times $ t_i $ which maximise information gain. As we shall see next, this speeds up the convergence of the temperature estimate and, more importantly, it largely reduces the spread of estimates drawn from small measurement records.

\subsubsection{\textit{A priori} optimisation of the expansion time} \label{sec:a-priori}

Even in the absence of measured data and starting from maximum ignorance about $ T $, one can make the most of the available information---namely, the parameters of the trap---in order to determine an optimal expansion time $ t_s $ for the first shot of the protocol. 
This amounts to searching for the time at which $\mathcal{K}(t)$ in Eq.~\eqref{eq:bayes-info} is maximal, which can be calculated \textit{before} any measurement is carried out. 
The simplest strategy would then consist in performing every subsequent measurement at the same recapture time $t_s$, i.e., 
\begin{equation*}
    \boldsymbol{t} = \smash[t]{(\underbrace{t_s, \dots, t_s}_\text{$\mu$ times})}.
\end{equation*}

The intuition behind this approach is that, by keeping the uncertainty of the most uninformed shot---the first one---to a minimum, one can boost the convergence of the estimate to the true temperature as knowledge is updated, as per Eq.~\eqref{eq:posterior-all-data}. 
This kind of \textit{a priori} optimisation was first explored in Mach--Zehnder interferometry \cite{jesus2018}, and it has been proven informative also in distributed sensing \cite{jesus2020mar}. For a discussion about the construction and application of single-shot optimised protocols, see Ref.~\cite{jesus2019thesis}.

\subsubsection{Adaptive optimisation of the expansion time}
\label{sec:adaptive}

Instead of continuing to measure at the \textit{a priori} optimal time $ t_s $, we can change the recapture intervals at each step so as to maximise the information content of the posterior probability \emph{adaptively} \cite{mehboudi2021fundamental}. Specifically, 
\begin{enumerate}
    \item Given the prior $p(\theta)$ and the likelihood $p(n|\theta, t)$ for the first shot, maximise  
    $\mathcal{K}(t)$ over $t$ to find $ t_1=t_s$. 
    \item Perform a measurement at $t_1 = t_s$ and record $n_1$. 
    \item Normalise $ p(\theta)\,p(n_1|\theta, t_1)$ and use it as the new `prior' for a second run \cite{jaynes2003, toussaint2011}.
    Then apply step 1 to find the optimal expansion time $t_2 $, and measure $n_2$. 
    \item Iterate $\mu$ times. The resulting data can then be processed using Eqs.~\eqref{eq:est} and \eqref{eq:error-bar-and-mle}.
\end{enumerate}

\subsection{Performance comparison on single-atom simulations}\label{sec:singl-atom}

Let us now make a preliminary analysis of how these Bayesian strategies compare to each other, and with the non-linear fit described in Sec.~\ref{sec:release--recapture}. To that end, we simulated a trap loaded deterministically with one atom, choosing a trap depth of $U_0/k_B = \SI{290}{\micro\kelvin}$ and a beam waist $w_0 = \SI{1.971}{\micro\metre}$. In these simulations, the temperature was set to $T= \SI{40}{\micro\kelvin}$. The parameters are taken to be the same as in our actual experiment.

For the non-linear fit and the unoptimised Bayesian protocol, we generated $30$ measurement outcomes at each of the seven expansion times $(5, 10, 20, 30, 50, 70, 100)\, \SI{}{\micro\second}$. 
The \textit{a priori} optimised and fully adaptive protocols were applied on simulated records with the same number of entries as the other two methods (i.e., $210$). 
All three Bayesian strategies used Jeffreys's prior in Eq.~\eqref{eq:prior}, with support between $\theta_{\mathrm{min}}=\SI{14.5}{\micro\kelvin}$ and $\theta_{\mathrm{max}}=\SI{125}{\micro\kelvin}$.
For these parameters, the \textit{a priori} optimal expansion time evaluates to $t_s = \SI{14}{\micro\second}$.

Fig.~\ref{fig1} illustrates the estimate variability of each method. The advantage of time optimisation seems clear, as both the \textit{a priori} and adaptively optimised methods (rhombs and squares, respectively) yield much more accurate estimates than the fits to data collected at unoptimised recapture times (circles and triangles). 
Furthermore, in the insets (a--c) we show how the time-optimised protocols converge to the true temperature with significantly less data than their unoptimised counterpart. In the next two sections we shall put these preliminary observations on a more solid basis. 

\section{Experimental realisation}\label{sec:real-experimental}

\subsection{The experimental setup}\label{sec:setup}

We now apply our Bayesian framework on actual experimental data. Details on our experimental setup can be found in \cite{mellado2020}. In this work, \ce{^{41}K} atoms are loaded into an optical tweezer from a dark magneto-optical trap. The number of atoms transferred into the tweezer can be controlled by varying the magneto-optical trap loading time. The tweezer is realised by focusing  a beam at $ \SI{790}{\nano\metre}$, produced by a Ti:sapphire laser, to a waist of $\simeq \SI{1.9}{\micro\metre}$ using a 20$\times$ apoplanar infinity corrected objective. This realises an elongated tweezer trap, with a radial-to-axial dimension ratio of $ 1\colon 249 $. As discussed in Sec.~\ref{sec:release--recapture}, such setup can be described by the model in Eq.~\eqref{eq:recapture} to a very good approximation. The temperature of the few-atom system is determined by the depth of the optical tweezer. Evaporation cooling brings down the temperature from its initial higher value of the atom source.

The release-recapture protocol is realised by switching off the tweezer's light for a variable amount of time and then switching it on again. To measure the number of atoms left in the tweezer, we utilise fluorescence imaging switching on a series of dedicated resonant beams. The light emitted by the recaptured atoms is collected with the same microscope objective that is used to create the tweezer, and detected with a sCMOS camera. 
In order to avoid anti-trapping while imaging \cite{Cheuk_Quantum_2015}, we chop the tweezer and imaging lights out of phase at $\SI{1}{\mega\hertz}$ for the duration of the measurement. Once the detection is completed, the \ce{^{41}K} atoms are lost. A new batch of atoms needs to be loaded in the tweezer, and the process above is repeated. 

\subsection{Mapping photon counts into atom numbers}

The experimental measurement record is a list of photon counts $ n_{p,i} $ at each shot at time $ t_i $. We must, however, convert each photon count into a number of recaptured atoms $ n_i $ to be fed into the vector $\boldsymbol{n}$. To do so, we take a set of calibration fluorescence images. The histogram of the resulting photon counts produces a multimodal distribution, where each peak is associated with a different number of atoms in the trap. Applying a multimodal Gaussian fit to these data gives us the position $ m $ of the peak corresponding to an empty trap, as well as the spacing $ \Delta $ between the (equispaced) peaks. Hence, the photon-count-to-atom conversion is simply
\begin{equation*}
    n_i = \textrm{round}\left(\frac{n_{p,i} - m}{\Delta}\right);
\end{equation*}
that is, we map each photon-count measurement to the atom number associated to the nearest peak. 

\subsection{Multi-atom model}
\label{sec:model}

Although the initial atom number can be controlled to some degree, our assumption of deterministic single-atom loading certainly needs to be relaxed before we can process experimental data. This can be achieved simply by upgrading the measurement model $p(n_i|T,t_i)$ in Eq.~\eqref{eq:model}. 

Assuming distinguishable atoms with negligible mutual interactions, the probability for recapturing $n_i$ of them during the $i$-th run, given that the trap was loaded with $N_0$ atoms, can be cast as 
\begin{subequations}\label{eq:multi-atom-model}
\begin{align} \label{eq:marginal-poisson}
p(n_i|N_0,T,t_i)= B[n_i|N_0, f(T,t_i)],    
\end{align}
where $B(n_i|N_0, q)$ is the binomial distribution
\begin{align}
    B(n_i|N_0,q) = \binom{N_0}{n_i} q^{n_i} (1-q)^{N_0 - n_i}
\end{align}
\end{subequations}
with success probability $q = f(T,t_i)$, as defined in Eq.~\eqref{eq:recapture}.

$N_0$ appears here as a nuisance parameter and thus needs to be marginalised over. To do so, we first note that, in this type of problem, the probability of starting with $N_0$ atoms can be described by a Poisson distribution \cite{ruschewitz1996investigations, haubrich1996observation}, denoted 
\begin{equation}\label{eq:poisson}
    P(N_0|\lambda) = \frac{\lambda^{N_0}}{N_0!}\,\mathrm{e}^{-\lambda},
\end{equation}
where $\lambda$ is the mean initial atom number.
In practice, this can be inferred from a sufficiently large set of calibration measurements at zero recapture time (i.e., on \textit{unreleased} atoms). 
In such case (cf. Appendix~\ref{app:poisson}),
\begin{align}\label{eq:poisson-binomial}
    p(n_i|T,t_i) &= \sum\nolimits_{N_0=0}^\infty P(N_0|\lambda)\,B[n_i|N_0, f(T,t_i)] 
    \nonumber \\
    &= P[n_i|\lambda f(T,t_i)].
\end{align}
The model for the multi-atom scenario is thus a Poisson distribution with mean equal to $\lambda f(T,t_i)$. With the sole exception of this change, the Bayesian protocols from Sec. \ref{sec:maximising-info} can be directly applied to experimental data.

\section{Results and discussion}
\label{sec:results}

\subsection{Processing real data at unoptimised recapture times}\label{sec:conventional-vs-Bayesian}

\begin{figure}[t]
\includegraphics[width=0.45\textwidth]{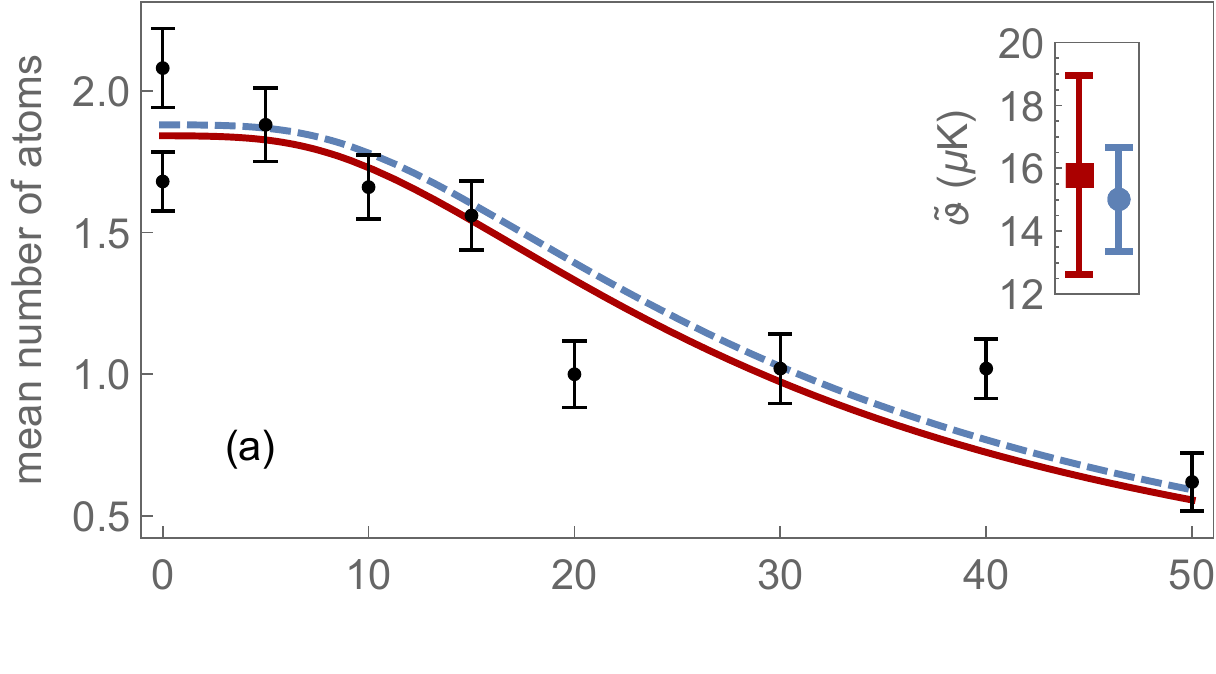}\vspace{-10pt}\\\includegraphics[width=0.45\textwidth]{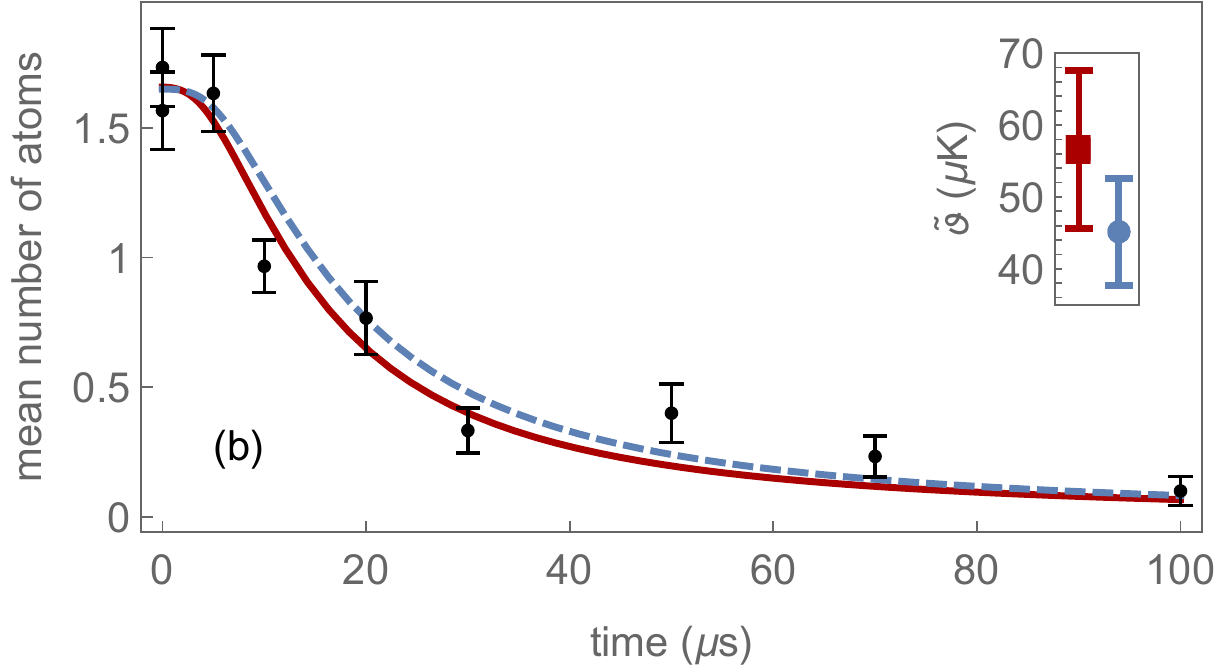}
\caption{\textbf{Temperature estimation from measurements at unoptimised recapture times.} 
Empirical mean and standard error of the number of recaptured \ce{^{41}K} atoms (black dots) at each free-expansion time in the two experiments described in Sec.~\ref{sec:conventional-vs-Bayesian} for a shallow (a) and deep trapping potential (b). The solid red curves are two-parameter non-linear fits to $ \lambda f(T,t) $ in $ \lambda $ and $ T $, where $ f(T,t) $ is the survival probability from Eq.~\eqref{eq:recapture}. These yield $T_a = \SI[separate-uncertainty=true]{15.8(32)}{\micro\kelvin}$ and $ \lambda_a = \SI[separate-uncertainty=true]{1.84(11)}{} $ (a), and $T_b = \SI[separate-uncertainty=true]{57(11)}{\micro\kelvin}$ and $ \lambda_b = \SI[separate-uncertainty=true]{1.66(9)}{} $ (b). In turn, the Bayesian strategy (cf. Sec.~\ref{sec:unoptimised-bayesian}) applied to the full measurement record gives $T'_a = \SI[separate-uncertainty=true]{15.0(17)}{\micro\kelvin}$ (a) and $T'_b = \SI[separate-uncertainty=true]{45(7)}{\micro\kelvin}$ (b). The insets show the estimates. The dashed blue curves correspond to $ \lambda f(t,T) $ evaluated at temperatures $ T_a' $ and $ T_b' $, with initial average atom numbers $ \lambda_a' = 1.88$ and $\lambda_b' = 1.65$, as calculated from the calibration measurements.}
\label{fig2}
\centering
\end{figure}

Let us start by analysing experimental measurement records taken at unoptimised light-off times. We shall apply the conventional method of least-squares fitting, as well as our Bayesian technique. Specifically, we shall work on two sets of measurements on \ce{^{41}K} atoms, which were performed in a shallow and a deep trapping potential. These had depth $U_0/k_B = \SI{110}{\micro\kelvin}$ and $\SI{290}{\micro\kelvin}$, respectively. For the shallow configuration, we performed $50$ measurements of the number of recaptured atoms at each of the expansion times $\boldsymbol{t}_a=(5, 10, 15, 20, 30, 40, 50)\,\SI{}{\micro\second}$, plus $100$ calibration measurements without release. 
In turn, for the deep trap, we took $30$ measurements at expansion times $\boldsymbol{t}_b=(5, 10, 20, 30,50,70,100)\, \SI{}{\micro\second}$, and $60$ calibration measurements at zero expansion time.

Let us now process these data via the non-linear fit to the mean number of recaptured atoms (cf. Sec.~\ref{sec:release--recapture}). We note that the average atom number at $ t=0 $ is also a point for the fit.
This returns temperature estimates of $\SI[separate-uncertainty=true]{15.8(32)}{\micro\kelvin}$ and $\SI[separate-uncertainty=true]{57(11)}{\micro\kelvin}$ for the shallow and deep configurations, respectively.
The fits are shown in Fig.~\ref{fig2} (solid red).

We next apply our Bayesian methods to the \emph{same} data.
From the calibration measurements, we find the initial mean atom numbers $\lambda=1.88$ and $\lambda=1.65$ for the shallow and the deep trap, respectively. As discussed in Sec.~\ref{sec:bayes-paradigm}, the support of our prior $ p(\theta) $ is set to $\SI{5.5}{\micro\kelvin} \leq \theta \leq \SI{30}{\micro\kelvin}$ (shallow) and $ \SI{14.5}{\micro\kelvin} \leq \theta\leq \SI{125}{\micro\kelvin}$ (deep).
Following the Bayesian procedure we arrive at the estimates $\SI[separate-uncertainty=true]{15.0(17)}{\micro\kelvin}$ and $\SI[separate-uncertainty=true]{45(7)}{\micro\kelvin}$ (dashed blue in Fig.~\ref{fig2}). 

Although the Bayesian and non-Bayesian estimates seem roughly comparable, we note that the Bayesian temperature estimate for the deep-trap experiment does fall below the error bars of the conventional method. Indeed, some discrepancy could have been expected in the second experiment, due to the smaller number of measurements performed. It is precisely in these cases when the Bayesian analysis pays off. 

\subsection{Finding the most reliable strategy}
\label{sec:res-a}

\begin{figure}[t]
\includegraphics[width=0.45\textwidth]{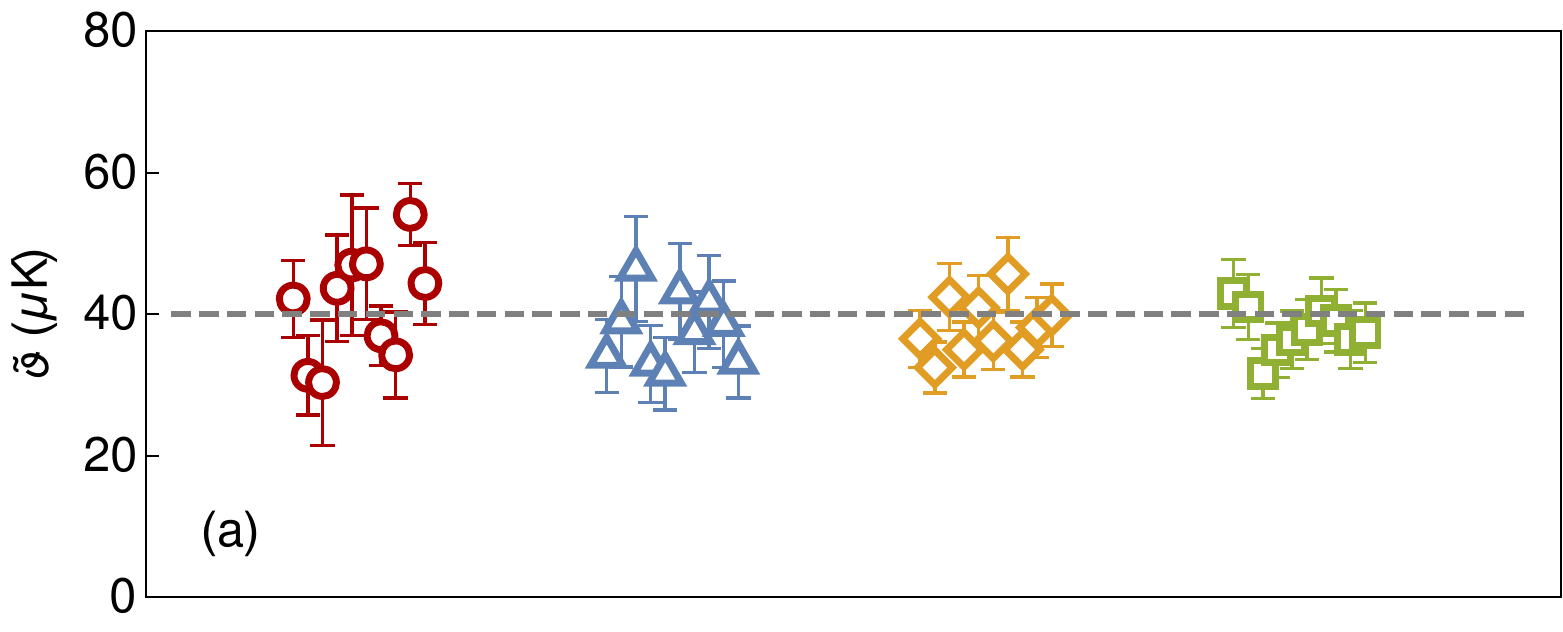}
\includegraphics[width=0.45\textwidth]{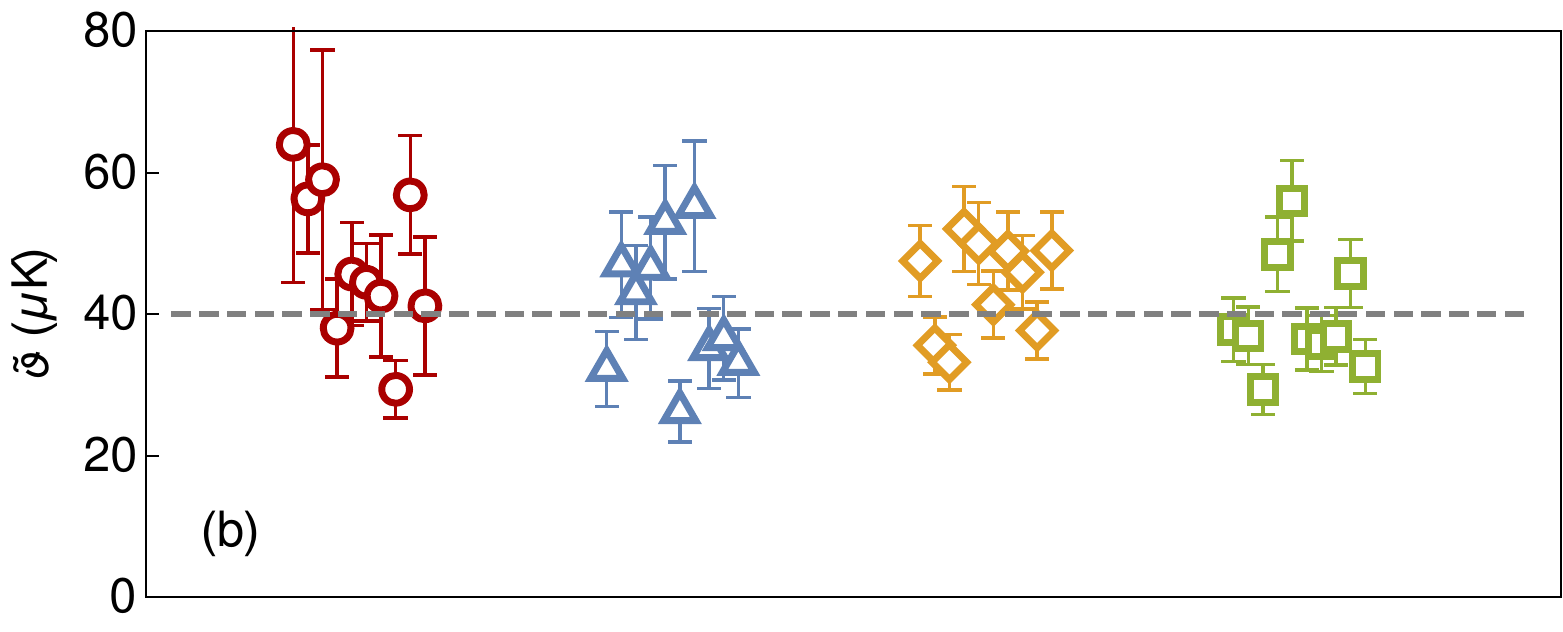}
\caption{\textbf{Estimate variability of different strategies.} Temperature estimates drawn from simulated release--recapture measurements on \ce{^{41}K}. The parameters for the simulations are those of the deep-trap experiment from Sec.~\ref{sec:conventional-vs-Bayesian} and Fig.~\hyperref[fig2]{2\,(b)}, and the true temperature is set to $\SI{40}{\micro\kelvin}$ (dashed grey). In (a), the mean number of atoms loaded into the trap is assumed to be known, and equal to $\lambda = 1.65$. In each case, $10$ independent runs of the experiment are simulated, consisting of $ 30 $ measurements generated at each of the $ 7 $ unoptimised recapture times $ \boldsymbol{t}_b $. The estimates are calculated via the conventional least-squares method (red circles) and the Bayesian approach (blue triangles). Since $\lambda$ is known, the least-squares method becomes a one-parameter fit. Orange rhombs and green squares correspond to estimates drawn from the \textit{a priori} optimised and fully adaptive protocols, respectively. These also run on datasets of $210$ measurements. The \textit{a priori} optimal recapture time is here $ t_s = \SI{22}{\micro\second}$. In (b) $\lambda$ is not known (its true value continues to be $\lambda = 1.65$). A calibration set of $60$ measurements at $t=0$ is thus needed, in addition to the $ 210 $ data points. In this case, the least-squares method becomes again a two-parameter fit in $T$ and $\lambda$. For all three Bayesian variants, $ \lambda $ is recalculated at every run from a fresh calibration set. Hence,  the $ \lambda $ used for estimation differs slightly between independent simulations. The overall variability of the estimates thus increases. This is particularly apparent for both the \textit{a priori} optimised and adaptive protocol, as the optimisations are done with the estimated  $ \lambda $, resulting in slightly off optimal times. Markers in (b) are the same as in (a).}
\label{fig3}
\centering
\end{figure}

Since the analysis above has shown that conventional and Bayesian processing of the same experimental data can yield different temperature estimates, we need to establish which one is more dependable and should thus be trusted. To that end, we now simulate multiple runs of the deep-trap experiment (see Fig.~\ref{fig3} for details), setting the temperature to $ \SI{40}{\micro\kelvin} $. 

Let us start by focusing on the red circles and blue triangles in Fig.~\hyperref[fig3]{3\,(b)}, which correspond to ten simulated deep-trap experiments processed with the conventional and Bayesian methods, respectively. Each run thus consists of $ 30 $ data points simulated at each of the $ 7 $ recapture times in $ \boldsymbol{t}_b $, plus $ 60 $ calibration measurements, at $t=0$. As we can see, the estimate variability is substantial, although the Bayesian estimates have a narrower spread and more consistent error bars. 

To quantify such variability, we use the empirical error \footnote{
Here we ignore the bias $(\langle\vartheta\rangle - T)$ with respect to the true temperature \cite{kay1993, rafal2015} because we are primarily concerned with the \textit{variability} of the estimate rather than its \textit{accuracy}. Nevertheless, we have verified that our conclusions remain valid even if a simulated bias term is included, i.e., $[\textrm{Var}\,\tilde{\vartheta} + (\langle \tilde{\vartheta}\rangle - T)^2]/\langle \tilde{\vartheta} \rangle^2$, with $T = \SI{40}{\micro\kelvin}$. As expected, this extra term vanishes asymptotically regardless of the estimation technique employed. Furthermore, in the limit of few data, the bias term has been found to be considerably smaller for Bayesian estimates. 
}. $\textrm{Var}\,\tilde{\vartheta}/\langle \tilde{\vartheta} \rangle^2$ of a set of estimates, where
\begin{subequations}\label{eq:variability}
\begin{align}
    \textrm{Var}\,\tilde{\vartheta} &= \langle \tilde{\vartheta}^2 \rangle - \langle \tilde{\vartheta} \rangle^2, \\
    \langle \tilde{\vartheta} \rangle &= \frac1N \sum\nolimits_i^N \tilde{\vartheta}_i.
\end{align}
\end{subequations}
For these parameters, and considering $100$ simulated runs, the variability of the conventional least-squares method evaluated to $ 0.064 $, while the Bayesian analysis proved slightly more reliable, with $ \textrm{Var}\,\tilde{\vartheta}/\langle \tilde{\vartheta} \rangle^2 = 0.057 $. 

However, the full potential of the Bayesian method is unleashed when the light-off times are picked to maximise information gain. Back to Fig.~\hyperref[fig3]{3\,(b)}, we now turn our attention to the results for the \textit{a priori} optimised and fully adaptive protocols (orange rhombs and green squares, respectively). These use datasets of the same length as the unoptimised schemes (i.e., $210$ points and $ 60 $ calibration measurements). In order to evaluate the information gain in Eq.~\eqref{eq:bayes-info}, we must truncate the Poisson distribution \eqref{eq:poisson} for the number of atoms initially loaded into the trap. Specifically, we set $ N_0 = 7 $ as the maximum, since the probability of loading more atoms is negligible for $ \lambda = 1.65 $ which is the true value of the initial number of atoms in our simulations. It is important to emphasise that $ \lambda $ is assumed unknown when processing data and thus, needs to be re-calculated from the calibration data at every simulated run (see caption of Fig.~\ref{fig3}). Furthermore, in order to account for the finite resolution when adjusting the recapture times in the lab, in our calculations we replace the exact optimal times by integer multiples of a minimum resolution of \SI{2}{\micro\second}.   

As it could be expected in light of Fig.~\ref{fig1}, the fully adaptive protocol is the most dependable, visibly outperforming its unoptimised counterparts. In terms of variability, $ \textrm{Var } \tilde{\vartheta}/\langle \tilde{\vartheta} \rangle^2 = 0.034 $ for $ 100 $ runs. That is a $ 40\% $ improvement with respect to the Bayesian processing of unoptimised data, and over $50\%$, with respect to the conventional method. Interestingly, the \textit{a priori} optimised protocol is comparable in variability to the fully adaptive one and yet, it is much simpler to implement. This is our first main result, and it supports using the \textit{a priori} optimised protocol as the new default when it comes to practical release--recapture thermometry. 

The fact that the initial average number of atoms in the tweezer is unknown certainly interferes with estimate variability. Namely, the empirical average of the simulated calibration data differs slightly from run to run, even if $ \lambda $ is fixed in the simulation. In order to single out the temperature-estimation aspect of the calculations, we also run variability checks, treating $ \lambda $ as a \textit{known} parameter. As we can see in Fig.~\hyperref[fig3]{3\,(a)}, the overall variability is much smaller in this case, but the hierarchy among the protocols is maintained. 

As an alternative, we might want to adapt our Bayesian--global processing to a multi-parameter scenario, so that both $ \lambda $ and $ T $ can be estimated. The procedure is summarised in Appendix~\ref{app:multi-parameter}. Although this certainly deserves further investigation, Bayesian multi-parameter adaptive estimation seems, at this stage, too time-consuming to be practical. 

\subsection{Assessing convergence speed}

\begin{figure}[t]
\includegraphics[width=0.45\textwidth]{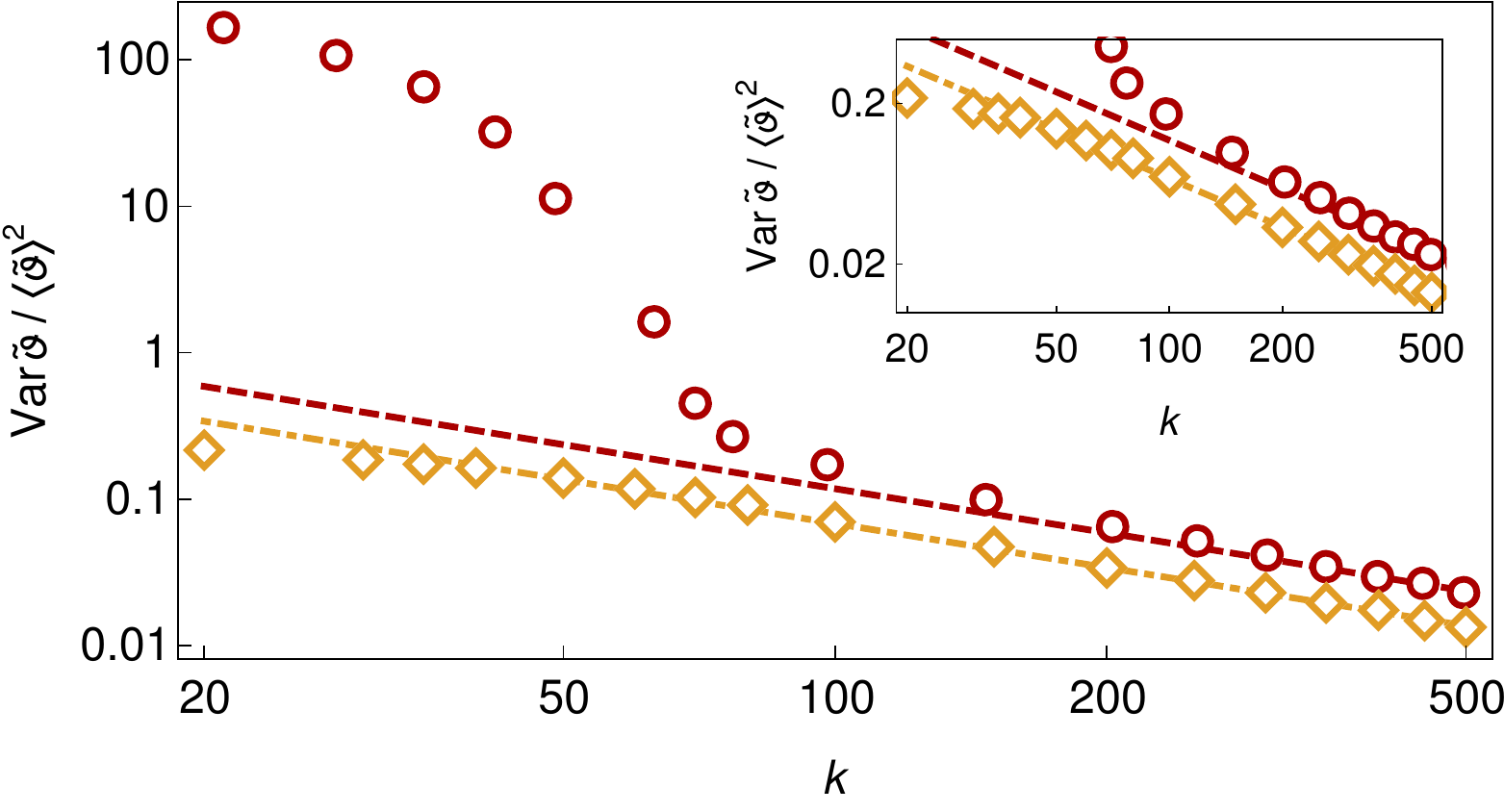}
\caption{\textbf{Convergence of conventional and \textit{a priori} optimised protocols.} Log--log plot of variability versus number of measurements $ k $, calculated from $5000$ simulated runs of the deep-trap experiment from Sec.~\ref{sec:res-a}. The measurement count $ k $ excludes any calibration measurements at $ t=0 $. Specifically, $(2/7)\,k$ calibration outcomes were generated for each run. All other parameters are the same as in Fig.~\ref{fig3}. The conventional non-linear-fit release--recapture (red circles) is compared to our \textit{a priori} optimised protocol (orange rhombs). Note the vast superiority of the optimised method for low $ k $, i.e., scarce data. Asymptotically, the variability takes the form $\sim 1/(kF)$. This occurs for $k > 350$ for the conventional protocol  and $k > 200$ for the optimised one. The dashed and dot-dashed lines highlight the asymptotic variability scaling for both methods. The offset between these lines indicates that the asymptotic convergence speed to the true temperature is roughly twice as fast for the \textit{a priori} optimised protocol. The insets zooms into the region in which our Bayesian technique enters the asymptotic regime.}
\label{fig4}
\centering
\end{figure}

So far we have shown that taking release--recapture data at the fixed \textit{a priori} optimal expansion time $ t_s $, and processing them with our global thermometry technique yields a clear advantage---in terms of reliability---over the conventional unoptimised protocol. However, in order to optimise the experimental resources it is also important to consider the `speed of convergence'; that is, the amount of data needed for the variability to settle into an asymptotic regime, and the rate at which the variability is reduced once in that limit.

In Fig.~\ref{fig4} we compare the conventional non-linear-fit method and our \textit{a priori} optimised Bayesian--global protocol in terms of convergence. We do this using the same parameters as in Sec.~\ref{sec:res-a}. Specifically, we compute the estimate variability from Eq.~\eqref{eq:variability} for $5000$ simulated runs of the experiment, and gradually increase the number of measurements per run $ k $. As it can be seen in the figure, as $ k $ grows the variability of both protocols eventually scales as
\begin{equation*}
    {\frac{\textrm{Var}\,\tilde{\vartheta}}{\langle \tilde{\vartheta} \rangle^2}} \sim \frac{1}{k F},
\end{equation*}
where $F$ is a free parameter \footnote{
The parameter $ F $ does not necessarily coincide with the Fisher information. $F$ would indeed stand for the Fisher information associated with $T$ if $\lambda$ were known exactly. 
However, here we are dealing with a multi-parameter problem, as both $T$ and $\lambda$ need to be determined. 
In principle, to guarantee optimality one would perform a two-parameter Bayesian estimation (see Appendix~\ref{app:multi-parameter}), which would further allow for the Cram\'{e}r--Rao bound to be saturated in the limit of many measurements \cite{rafal2015}. 
In that case, $F$ would correspond to the diagonal element of the inverse of the Fisher information matrix that is associated with temperature, thus involving a combination of the off-diagonal elements of such a matrix \cite{kay1993}. 
Yet, in this work, $\lambda$ is separately estimated by taking the empirical mean on unreleased atom numbers. Therefore, our overall estimator needs not to be efficient and the Cram\'er--Rao bound may not be saturated.}.
To get a rough idea of the onset of this scaling, we take the logarithm of the calculated variabilities and fit it to $ -a\log{k} - \log{F} $. We start from the largest $ k $ and work backwards, adding points at lower $ k $ to the fit. We then search for the number of shots at which the exponent $ a $ first deviates from $ 1 $ by more than $ 2.5\% $. This gives $ k_\text{c} \simeq 350  $ for the coventional method and $ k_\text{ap}\simeq 200 $ for our \textit{a priori} optimised protocol.

Hence, for $ k > k_\text{c}$ we can be certain that both protocols have entered the asymptotic regime in $ k $. From their ratio we find that the \textit{a priori} optimised method affords a $ 43\% $ variability reduction with respect to the common practice in release--recapture, even in the asymptotic limit. We have thus shown that our method is not only more reliable when processing short data sets, but that it also requires less measurements to attain the same precision as the conventional protocol. In our experimental setup, this roughly translates into \textit{halving} the number of measurements needed to hit any target precision. This is our second main result.

\subsection{Time optimisation on an individual experiment}

\begin{figure}[t]
\includegraphics[width=0.45\textwidth]{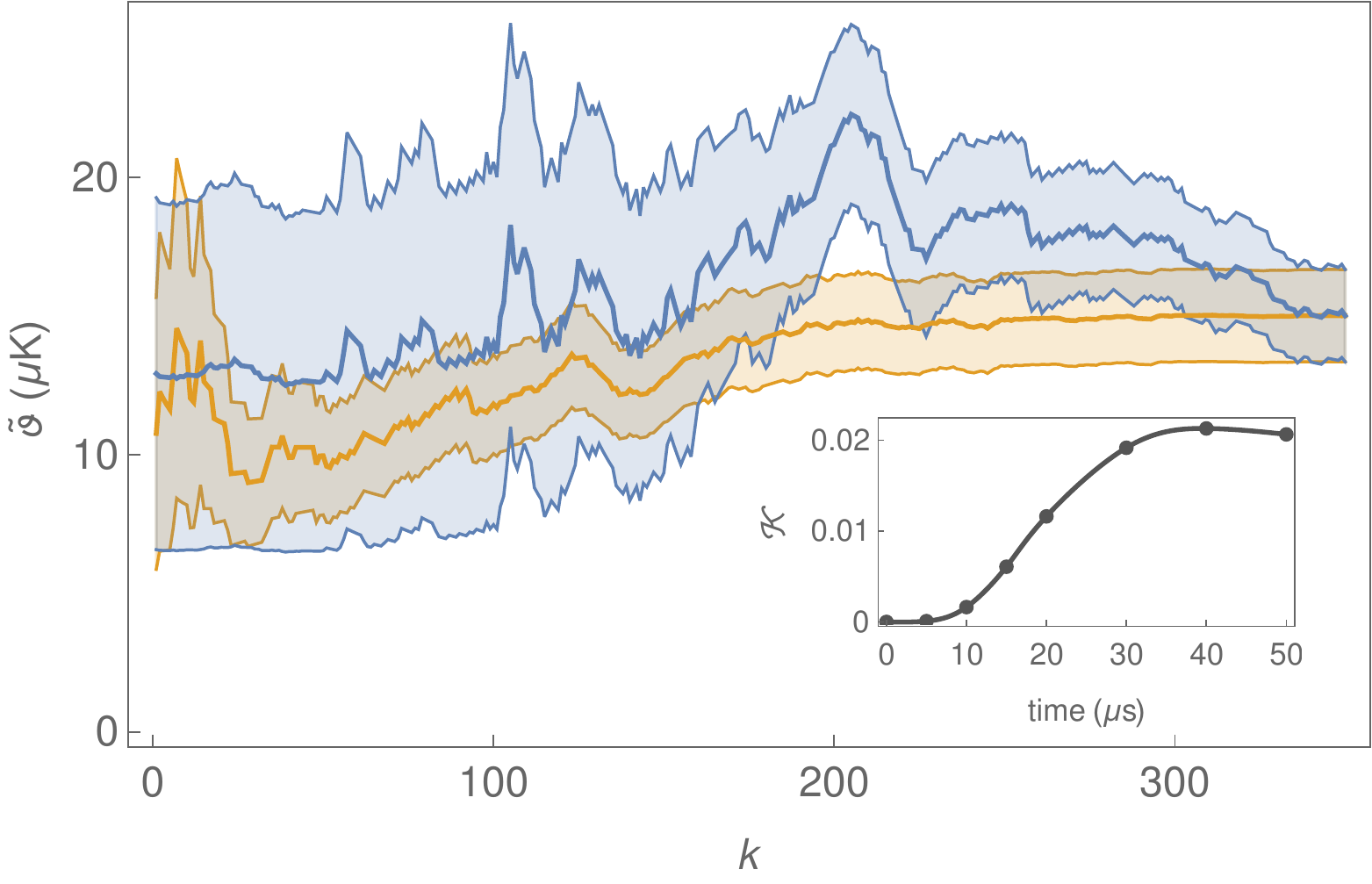}
\caption{\textbf{Recapture-time optimisation.} Estimate and error bar from the first $ k $ entries of the measurement record of the shallow-trap experiment from Fig.~\hyperref[fig2]{2\,(a)}, reordered to mimic the \textit{a priori} optimised protocol (orange). The optimal recapture time $ t_s=\SI{42}{\micro\second} $ was calculated as per Sec.~\ref{sec:a-priori}. The \textit{a priori} information gain $\mathcal{K}$ as a function of the recapture time is shown in the inset for this model (grey dots). The gray line is a mere guide to the eye. The $50$ measurements at the closest recorded time ($ \SI{40}{\micro\second} $) were moved to the top of the list. The set of measurements at the second closest recorded time ($ \SI{50}{\micro\second} $) were picked next, and so forth, until exhausting the $350$ available data. The estimate and error bars of that exact same list, but processed in the reverse order are shown in blue, for comparison. As we can see, the orange curve already converges to the final estimate by processing the $200$ most informative data. At that point, the estimate is $\SI[separate-uncertainty=true]{14.7\pm 1.7}{\micro\kelvin}$. The remaining $ 150 $ data carry practically no information, since the estimate drawn from the full measurement record is $ \SI[separate-uncertainty=true]{15\pm 1.7}{\micro\kelvin} $.}
\label{fig5}
\centering
\end{figure}

We now illustrate what an \textit{a priori} optimised release--recapture experiment might look like (cf. Fig.~\ref{fig5}). To that end, we retrospectively analyse the data from the shallow-trap experiment processed in Fig.~\hyperref[fig2]{2\,(a)}. Just like in Sec.~\ref{sec:res-a}, we cap the distribution of initial atom numbers in the trap at $N_0 = 7$, which allows us to compute $ t_s=\SI{42}{\micro\second} $ (see inset in Fig.~\ref{fig5}). 
We then process, with our Bayesian--global technique, those measurement outcomes captured at the times closest to $ t_s $. In this case, those would be the $ 50 $ measurements at $ \SI{40}{\micro\second} $. The a priori optimised method from Sec.~\ref{sec:a-priori} would demand to continue measuring at the same time. However, having run out of data, we move on to process the measurements at the second closest time in the dataset (i.e., $ \SI{50}{\micro\second} $), and so on.

The estimate and error bars resulting from the Bayesian processing of the first $ k $ entries of the reordered record are plotted in orange in Fig.~\ref{fig5}. We observe that over $ 100 $ points seemingly add no information about temperature, and could thus have been spared. That is, by aiming to uniformly sample the decay of the survival probability, nearly $ 30\% $ of the experimental resources were wasted. To further emphasise the important difference that time-optimisation can make to the convergence of the estimate, we overlay (in blue) a similar plot, in which the measurement record was ordered from lowest to largest information gain per entry.

Hence, the effect of optimising over the recapture time to increase information gain is threefold. First, it produces more reliable estimates when data is scarce. Secondly, it ensures a faster convergence towards the true temperature in the asymptotic limit of many measurements. Finally, it allows to significantly reduce the number of data points needed for the point estimate to converge (i.e. to stabilise at a certain value) in a \textit{single} experimental run. The latter observation is the final main result of this work. 

\subsection{\textit{A priori} optimised versus fully adaptive protocol}\label{sec:showdown}

\begin{figure}[t]
\includegraphics[width=0.45\textwidth]{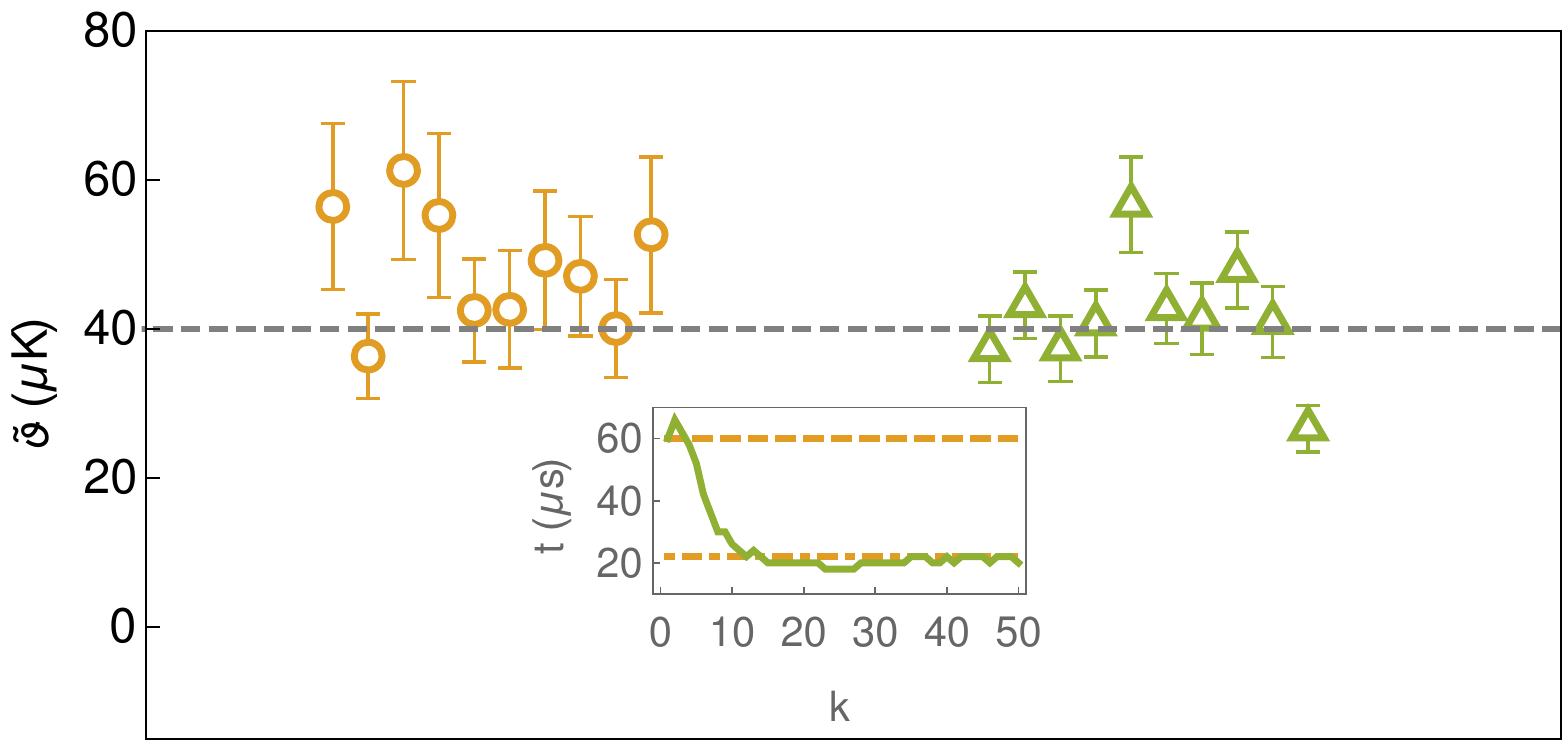}
\caption{\textbf{\textit{A priori} optimised vs. adaptive protocol.} Comparison between the \textit{a priori} optimised (orange rhombs) and fully adaptive (green squares) protocols. Ten independent runs of the deep-trap experiment were simulated for each method. All parameters are the same as in Fig.~\hyperref[fig3]{3\,(b)}, except for $ \theta_\text{min} $, which is set here to $ 0.5\% $ of the trap depth $ U_0/k_B $ (see Sec.~\ref{sec:showdown}). The inset shows the optimal expansion time as a function of $k$ for one of the simulated measurement records (solid green). The \textit{a priori} optimal recapture time for this wider hypothesis support is indicated by the orange dashed line. In contrast, the \textit{a priori} optimal time for the narrower support used in Figs.\ref{fig1}--\ref{fig5} is the orange dot-dashed line. Our observation that the \textit{a priori} optimised method seems to perform equally well than the arguably superior fully adaptive scheme is explained by the fact that, for the narrower prior, $ t_s $ almost coincides with the true optimum for most of the experiment.}
\label{fig6}
\centering
\end{figure}

As illustrated in Sec.~\ref{sec:res-a}, the \textit{a priori} optimised and the adaptive protocols can have almost identical estimate variability. This advised to focus on the former, due to its simplicity. In some situations, however, the effort of adaptively recalculating optimal expansions in real time might pay off. To see this, we go back to the deep-trap configuration of Fig.~\hyperref[fig2]{2\,(b)}, to make another showdown between these two protocols. Proceeding exactly as in Sec.~\ref{sec:res-a}, we simulate $ 10 $ independent experiments for each method but, in this case, we work with a wider support for our prior, stretching it by one order of magnitude into colder temperatures. Namely, $\SI{1.45}{\micro\kelvin}\leq\theta\leq\SI{125}{\micro\kelvin}$. 

As we can see in Fig.~\ref{fig6}, the adaptive protocol exhibits here a reduced variability with respect to the \textit{a priori} optimised method. This is due to the fact that the \textit{a priori} optimal time is nearly $ \SI{60}{\micro\second} $, while the true optimal release--recapture delay stabilises around $ \SI{20}{\micro\second} $ after few measurements (see inset in Fig.~\ref{fig6}). This quickly renders the \textit{a priori} optimised method inefficient when it comes to information acquisition. The fully adaptive protocol, however, does not share this problem.

\section{Conclusions}
\label{sec:conclusion}
In this paper we have demonstrated how temperature measurements can be substantially improved in practice by actively exploiting Bayesian--global thermometry techniques. We do so on a concrete experimental setup; namely, release--recapture measurements on few tightly confined \ce{^{41}K} atoms at \SI{}{\micro\kelvin} temperatures. The conventional protocol samples the fraction of recaptured atoms evenly and fits the data to a model for the recapture probability. Instead, we propose to choose the delay between release and recapture so as to maximise information gain, and to process the measured data with our Bayesian technique \cite{rubio2021global}.

Working with real experimental data as well numerical simulations, we have shown that our method has three important advantages. 
First, the variability of estimates extracted from small data sets is largely reduced when compared to the conventional protocol. This results in much more dependable thermometry. Second, time optimisation also leads to a faster convergence towards the true temperature in the asymptotic limit of many measurements. Finally, the optimisation of the recapture time substantially reduces the number of measurements needed for estimates to stabilise to a final temperature reading. Put differently, the conventional approach wastes resources by performing and processing a sizeable proportion of uninformative measurements.

We have also demonstrated that the best results are obtained with a fully adaptive approach, in which each new release--recapture interval takes in all previous measurements. Interestingly, a simpler protocol can yield surprisingly competitive estimates; namely, measuring repeatedly at a single expansion time---the \textit{a priori} optimal one. That is, the delay that maximises the information gain in the first measurement. This follows from the parameters of the experiment and the range of temperatures being probed. Other than an arbitrarily broad temperature range, we have assumed no prior knowledge about the temperature.

We have shown how quantum thermometry can deliver practical solutions leading to quantifiable precision enhancements, once the quantum-Fisher-information based `local' paradigm \cite{mehboudi2019review} is abandoned in favour of the global framework. While the former has proven useful when studying fundamental precision limits and maximising the \textit{responsiveness} of temperature sensors, the more general Bayesian framework can, in addition to that \cite{demkowicz2020,rubio2021scales}, process finite experimental records optimally. Importantly, the exact same fundamental principles can be applied to other techniques in different temperature ranges and experimental platforms, which opens a new exciting avenue in quantum thermometry.

\begin{acknowledgments}

We thank Robert Smith for useful comments.
J.G. is funded by the College of Engineering, Mathematics and Physical Sciences of the University of Exeter. 
J.R. acknowledges support from EPSRC (Grants No. EP/T002875/1 and EP/R045577/1). 
The work of R.S., T.H. and G.B. was supported by the Leverhulme Trust Research Project Grant UltraQuTe (Grant No. RGP-2018-266).

\end{acknowledgments}

\appendix

\section{Scale-invariant thermometry}\label{sec:scale-est-theory}

This appendix applies the framework of (quantum) scale estimation in Refs.~\cite{rubio2021global, rubio2021scales} to release-recapture thermometry. 

\subsection{Temperature as a scale parameter}\label{app:scale-parameter}

Let us first examine the multi-atom measurement model for our  experiment. That is, the probability function $p(n_i | T, t_i)$ for recapturing $n_i$ atoms after time $t_i$ during the $i$-th run, given that their temperature is $T$.
As shown in Secs.~\ref{sec:release--recapture} and \ref{sec:model} and in Appendix~\ref{app:survival}, $p(n_i | T, t_i)$ takes the form
\begin{equation}
    p(n_i | T, t_i) = \frac{\mathrm{e}^{-\lambda f(T, t_i)}}{n_i!}[\lambda f(T, t_i)]^{n_i}.
\end{equation}
Since both temperature and the other dimensioned quantities appear only within the fraction of recaptured atoms, we can base our analysis on the function $f(T, t)$ alone. 
This implies that the discussion in this appendix also applies to the single-atom model in the main text. 

Let us now define the kinetic energy
\begin{equation}
    E_k \coloneqq \frac{1}{2} m \left( \frac{l}{t} \right)^2
\end{equation}
of an atom having moved by the characteristic length scale $l = w_0/\sqrt{2}$ at time $t$.
Using this, the fraction of recaptured atoms can be cast as
\begin{equation}
    f(T,t) = \frac{g\lbrace[E_k/(k_B T)]\, W(U_0/E_k)\rbrace}{g[U_0/(k_B T)]}.
\end{equation}
That is, our statistics is invariant under transformations $U_0' \mapsto \gamma U_0$, $E_k' \mapsto \gamma E_k$ and $T' \mapsto \gamma T$, where $\gamma$ is an arbitrary positive constant.   
This implies that temperature sets the energy scale of the problem, and so it is \emph{scale parameter} from the point of view of estimation theory \cite{jaynes1968, prosper1993, jaynes2003, rubio2021global, rubio2021scales}. 

\subsection{Protocol optimisation}

This work optimises release--recapture experiments in two ways. 
First, by constructing an estimator function mapping the measurement outcomes $\boldsymbol{n} = (n_1, \dots, n_\mu)$ and the release times $\boldsymbol{t} = (t_1, \dots, t_\mu)$ to the \emph{optimal} temperature estimate $\tilde{\vartheta}(\boldsymbol{n}, \boldsymbol{t})$, where $\mu$ is the number of experimental runs.
Secondly, by identifying the most informative values for the release times $\boldsymbol{t} = (t_1, \dots, t_\mu)$. 

Given that $T$ is unknown, both optimisations need be carried out with respect to a temperature-independent measure of uncertainty. 
In addition, they need be performed \emph{before} the experiment is run (see Sec.~\ref{sec:maximising-info}), meaning that the uncertainty must also be outcome-independent. 
We thus use the uncertainty quantifier 
\begin{equation}
    \bar{\epsilon}(\boldsymbol{t}) = \int d\boldsymbol{n}\,p(\boldsymbol{n}|\boldsymbol{t})\,\bar{\epsilon}(\boldsymbol{n}, \boldsymbol{t}),
    \label{eq:general-err}
\end{equation}
where
\begin{equation}
    \bar{\epsilon}(\boldsymbol{n},\boldsymbol{t}) = \int d\theta\,p(\theta|\boldsymbol{n},\boldsymbol{t})\,\mathcal{D} [ \tilde{\theta}(\boldsymbol{n}, \boldsymbol{t}), \theta ].
    \label{eq:exp-err-app}
\end{equation}
Here, $\theta$ is a hypothesis about the true value of $T$, $\tilde{\theta}(\boldsymbol{n}, \boldsymbol{t})$ is a generic temperature estimator, and $\mathcal{D} [ \tilde{\theta}(\boldsymbol{n}, \boldsymbol{t}), \theta ]$ is a deviation function gauging the deviation of $\tilde{\theta}(\boldsymbol{n}, \boldsymbol{t})$ from $\theta$. 
Furthermore, $p(\boldsymbol{n}| \boldsymbol{t}) = \int d\theta\,p(\theta)\,p(\boldsymbol{n}|\theta,\boldsymbol{t})$, while the probabilities $p(\theta)$, $p(\boldsymbol{n}|\theta,\boldsymbol{t})$ and $p(\theta|\boldsymbol{n},\boldsymbol{t})$ are defined in the main text and in Appendix~\ref{app:posterior-derivation}. 

To use Eqs.~\eqref{eq:general-err} and \eqref{eq:exp-err-app}, we must choose a deviation function $\mathcal{D}$. 
As shown in Ref.~\cite{rubio2021global}, the scale-invariant nature of the protocols in this work (see Sec.~\ref{app:scale-parameter}) leads to the logarithimic deviation function $\mathcal{D} [ \tilde{\theta}(\boldsymbol{n}, \boldsymbol{t}), \theta ] = \log^2[\tilde{\theta}(\boldsymbol{n}, \boldsymbol{t})/\theta]$. 
Using this, Eqs.~\eqref{eq:general-err} and \eqref{eq:exp-err-app} become the mean logarithmic errors $\bar{\epsilon}(\boldsymbol{t}) \mapsto \bar{\epsilon}_{\mathrm{mle}}(\boldsymbol{t})$ and
\begin{equation}
    \bar{\epsilon}(\boldsymbol{n},\boldsymbol{t}) \mapsto \bar{\epsilon}_{\mathrm{mle}}(\boldsymbol{n},\boldsymbol{t}) = \int d\theta\,p(\theta|\boldsymbol{n},\boldsymbol{t}) \log^2{\left[\frac{\tilde{\vartheta}(\boldsymbol{n},\boldsymbol{t})}{\theta}\right]},
    \label{eq:mle-exp-appendix}
\end{equation}
respectively. 
Unlike the more familiar square errors, which are based on the deviation function $[\tilde{\theta}(\boldsymbol{n}, \boldsymbol{t}) - \theta]^2$ \cite{jaynes2003, rafal2015}, logarithmic errors respect the scale invariance of the problem, i.e., $\mathcal{D}(\gamma \tilde{\theta}, \gamma \theta) = \mathcal{D}(\theta, \tilde{\theta})$, and they do so while relying on minimal assumptions \cite{rubio2021global, rubio2021scales}; 
hence, their use in this work. 
We can then minimise $\bar{\epsilon}_{\mathrm{mle}}(\boldsymbol{t})$ with respect to $\tilde{\theta}(\boldsymbol{n}, \boldsymbol{t})$, finding the optimal temperature estimator \cite{rubio2021global}
\begin{equation}
    \tilde{\vartheta}(\boldsymbol{n},\boldsymbol{t}) = \theta_u \exp\left[\int d\theta \,p(\theta|\boldsymbol{n},\boldsymbol{t}) \log{\left(\frac{\theta}{\theta_u}\right)}\right]
    \label{eq:est-app}
\end{equation}
used in the main text.

To find the optimal release times $\boldsymbol{t}$, we first insert the estimator above into the expression for $\bar{\epsilon}_{\mathrm{mle}}(\boldsymbol{t})$, arriving at the fundamental lower bound $\bar{\epsilon}_{\mathrm{mle}}(\boldsymbol{t}) \geq \bar{\epsilon}_p - \mathcal{K}(\boldsymbol{t})$ \cite{rubio2021global, rubio2021scales}.
Such a bound is the result of taking the uncertainty prior to performing the experiment, denoted by $\bar{\epsilon}_p$, and subtracting the mean information gain $\mathcal{K}(\boldsymbol{t})$ associated with the measurement protocol under consideration. 
The expression for $\mathcal{K}(\boldsymbol{t})$ is given in the main text for a single shot, while that  for $\bar{\epsilon}_p$ can be found in Refs.~\cite{rubio2021global, rubio2021scales}. For our purposes, the key property of $\bar{\epsilon}_p$ is its lack of dependence on the release times $\boldsymbol{t}$. 
Given this, if we wish to optimise our protocol by minimising the bound above for a single shot, as done in this work, we simply need to replace $\boldsymbol{t} \mapsto t$ and maximise $\mathcal{K}(t)$ with respect to $t$. 

We finally note that, once the optimal strategy has been found, the theoretical error in Eq.~\eqref{eq:general-err} ceases to be relevant from an experimental point of view. 
Indeed, since our final goal is processing real data, the associated error bars must rely on a temperature-independent but outcome-dependant uncertainty, i.e., that in Eq.~\eqref{eq:mle-exp-appendix}.

\section{Posterior probability}
\label{app:posterior-derivation}

In this appendix we derive Eq.~\eqref{eq:posterior-all-data}. 
First, the posterior probability $p(\theta|\boldsymbol{n},\boldsymbol{t})$ is given by Bayes theorem as \cite{jaynes2003, toussaint2011} 
\begin{equation*}
    p(\theta|\boldsymbol{n},\boldsymbol{t}) \propto p(\theta|\boldsymbol{t}) p(\boldsymbol{n}|\theta, \boldsymbol{t}).
\end{equation*}

We now note that the expansion times $\boldsymbol{t} = (t_1, \dots, t_\mu)$ do not (by themselves) inform how likely different hypothesis $\theta$ about $T$ are. 
Hence, we take $p(\theta | \boldsymbol{t}) \mapsto p(\theta)$. 
Assuming statistical independence between measurements further leads to 
\begin{equation*}
    p(\boldsymbol{n}|\theta,\boldsymbol{t}) = \prod\nolimits_{i=1}^\mu p(n_i|\theta,\boldsymbol{t}).
\end{equation*}

But we also see that expansion times other than $t_i$ do not inform the likelihood for number of recaptured atoms $n_i$ recorded in the $i$-th measurement.
One thus finally arrives at 
\begin{equation*}
    p(\theta|\boldsymbol{n},\boldsymbol{t}) = \frac{1}{A}\,p(\theta) \prod\nolimits_{i=1}^\mu p(n_i|\theta,t_i),
\end{equation*}
as stated in the main text. The normalisation factor $ A $ is found by integrating over the hypothesis range $[\theta_{\mathrm{min}},\theta_{\mathrm{max}}]$, i.e.,
\begin{equation*}
    A = \int_{\theta_{\mathrm{min}}}^{\theta_{\mathrm{max}}} d\theta\,p(\theta) \prod\nolimits_{i=1}^\mu p(n_i|\theta,t_i).
\end{equation*}

\section{Calculation of the recaptured fraction}\label{app:survival}

For the sake of completeness, we summarise the main steps of the derivation of Eq.~\eqref{eq:recapture} as given in Ref.~\cite{mudrich203phd}. 
Let us load $ N_0 $ atoms of mass $m$ at low temperature, i.e., $k_B T \ll U_0 $, into the tweezer trap. Specifically, the temperature is low enough so that the atoms are well localised at the centre of the trap.
Furthermore, we shall neglect interatomic interactions, the influence of gravity, and the movement along the $z$-axis.
After a free expansion time  $ t $, the trap is re-activated. An atom may only be recaptured then if, 
\begin{equation*}
   \frac{1}{2}\,m \left(\frac{r}{t} \right)^2 \le U_0 \exp\left(- \frac{2 r^2}{w_0^2}\right),
\end{equation*}
where $ r $ is the distance from the center of the trap (see Sec.~\ref{sec:release--recapture}).
One then defines $ r_\text{max} $ as the distance which saturates the above inequality. It is convenient to introduce the dimensionless variables $\tilde{r}^2 = 2r^2/w_0^2$ and $\tilde{t}^2=4 U_0 t^2/(m w_0^2)$, so that $ \tilde{t}^2 = \tilde{r}^2_\mathrm{max}\exp(\tilde{r}^2_\mathrm{max}) $ or, equivalently, $ \tilde{r}^2_{\mathrm{max}} = W(\tilde{t}^2) $, where 
where $W(\cdots)$ is the product-logarithm or Lambert-$ W $ function defined in the main text.
Hence, $r_{\mathrm{max}}^2 = w_0^2\,W(\tilde{t}^2)/2$, so the maximal speed of a recaptured atom is
\begin{equation*}
    v^2_\textrm{max}(t) = \left(\frac{r_\textrm{max}}{t}\right)^2 =  w_0^2\,\frac{W(\tilde{t}^2)}{2 t^2}.
\end{equation*}

Only those particles with velocity below $ v_\text{max}(0) = \sqrt{2U_0/m} $ would be initially captured by the trap. On the other hand, only those with velocity under $ v_\text{max}(t) $ would be recaptured after free expansion during time $ t $. Therefore, the recapture fraction simply writes as
\begin{equation}
  f(T,t) = \frac{\int_0^{v_\textrm{max}(t)} dv\,F(v)}{\int_0^{v_\text{max}(0)} dv\,F(v)}
  = \frac{1}{g(\eta)}\,g{\left[\frac{\eta\,W(\tilde{t}^2)}{\tilde{t}^2}\right]},
  \label{eq:fraction-app}
\end{equation}
where $ F(v) = \frac{m v}{k_B T}\,\exp\left[-m v^2/(2 k_B T)\right]$ is the Maxwell--Boltzmann distribution in 2D,
$g(s) = 1 - \mathrm{e}^{-s}$, and $\eta=U_0/(k_B T)$.
Eq.~\eqref{eq:fraction-app} follows from
\begin{equation*}
\int_0^{u}dv\, F(v) =  g{\left(\frac{m u^2}{2 k_B T}\right)}.
\end{equation*}

Eq.~\eqref{eq:fraction-app} is useful due to its simplicity. 
However, we may loosen most of the underlying assumptions by simulating many individual trajectories of atoms. At low-enough temperatures, these can be sampled from a Gaussian distribution in position as well as in velocity in spite of the anharmonicity of the trap. The effects of gravity and the subsequent axial motion can also be included. 

\begin{figure*}[t]
\includegraphics[width=0.318\textwidth]{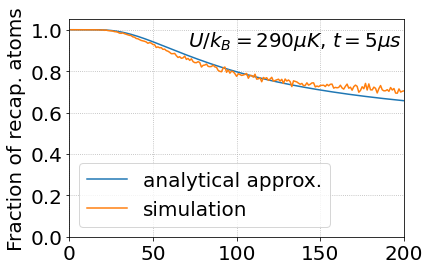}
\includegraphics[width=0.3\textwidth]{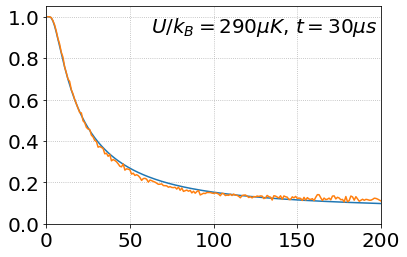}
\includegraphics[width=0.3\textwidth]{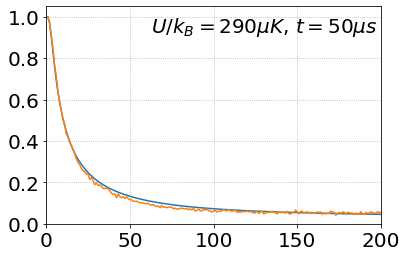}

\includegraphics[width=0.318\textwidth]{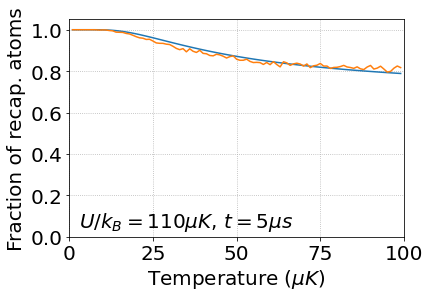}
\includegraphics[width=0.3\textwidth]{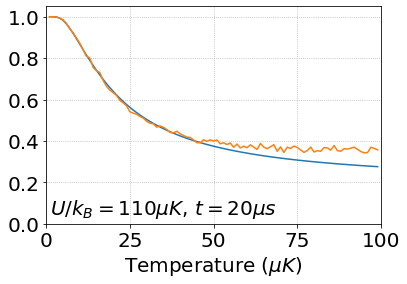}
\includegraphics[width=0.3\textwidth]{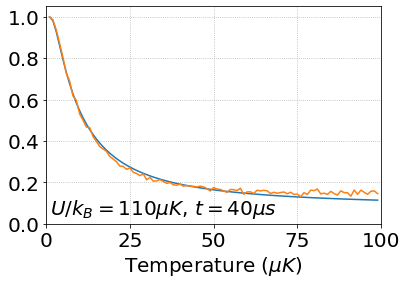}
\caption{\textbf{Regime of validity of Eq.~(\ref{eq:fraction-app}) for the recaptured fraction.} 
The analytical approximation Eq.~\eqref{eq:fraction-app} (blue) is compared to the recaptured fraction computed from $ 5000 $ individual simulated trajectories of atoms sampled from a thermal distribution (orange), as a function of temperature, for different recapture times. Axial motion due to gravity is also factored in the calculation.
As it can be seen, for a deep trap with $ U_0/k_B = \SI{290}{\micro\kelvin} $ (top row), the approximation holds well up to $ \sim \SI{125}{\micro\kelvin} $, while it breaks down at $ \sim \SI{30}{\micro\kelvin} $ in a shallow trap configuration (bottom row) with $ U_0/k_B = \SI{110}{\micro\kelvin} $. Simulations are for \ce{^{41}K} and the trap geometry described in Sec.~\ref{sec:setup}.}
\label{figA}
\centering
\end{figure*}

As we can see in Fig.~\ref{figA}, our simple model from Eq.~\eqref{eq:fraction-app} is in very good agreement with the numerically calculated recaptured fraction over a broad range of temperatures and recapture times. As expected, some deviations start to be appreciated as the temperature nears $ U_0/k_B $. 

It must be noted that limiting ourselves to the regime in which Eq.~\eqref{eq:recapture} is applicable is not as restrictive as it might seem. As we can see, the numerically calculated recaptured fraction becomes almost independent of $T$ when model and simulation start to diverge. Hence, when Eq.~\eqref{eq:recapture} ceases to apply, release--recapture thermometry itself becomes inefficient.

\section{Proof of Eq.~(\ref{eq:poisson-binomial}) in the main text}\label{app:poisson}

Starting from the left-hand side of Eq.~\eqref{eq:poisson-binomial}, we have
\begin{align*}
    p(n_i|T,t_i) &= \sum_{N_0=0}^\infty P(N_0|\lambda) B[n_i|N_0,q = f(T,t_i)] 
    \nonumber \\
    &= \sum_{N_0=0}^\infty \frac{\lambda^{N_0} e^{-\lambda}}{N_0!} \binom{N_0}{n_i} q^{n_i} (1-q)^{N_0 - n_i},
\end{align*}
where we recall that $q = f(T,t_i)$. 
Given that
\begin{equation*}
    \binom{N_0}{n_i} = 0
\end{equation*}
for $N_0 < n_i$, we only need to sum from $ N_0 = n_i$, i.e., 
\begin{align*}
    p(n_i|T,t_i) &= \sum_{N_0=n_i}^\infty \frac{\lambda^{N_0} e^{-\lambda}}{N_0!} \binom{N_0}{n_i} q^{n_i} (1-q)^{N_0 - n_i},
    \nonumber \\
    &= \sum_{N_0=n_i}^\infty \frac{\lambda^{N_0} e^{-\lambda}}{N_0!} \frac{N_0!}{n_i!(N_0 - n_i)!}
    q^{n_i} (1-q)^{N_0 - n_i}
    \nonumber \\
    &= \frac{q^{n_i} e^{-\lambda}}{n_i!}\sum_{N_0=n_i}^\infty \frac{\lambda^{N_0}}{(N_0-n_i)!} (1-q)^{N_0-n_i}. 
\end{align*} 
Next, we introduce the new variable $j = N_0 - n_i$, so that
\begin{align*}
    p(n_i|T,t_i) &= \frac{q^{n_i} e^{-\lambda}}{n_i!}\sum_{j=0}^\infty \frac{\lambda^{j+n_i} }{j!} (1-q)^{j}  
    \nonumber \\
    &=\frac{(\lambda q)^{n_i} e^{-\lambda}}{n_i!}\sum_{j=0}^\infty \frac{[\lambda (1-q)]^j }{j!}.
\end{align*} 
But the remaining sum is just the Taylor expansion of the exponential function $\mathrm{e}^{x}$ around $x=0$. 
Therefore, 
\begin{align}
    p(n_i|T,t_i) &= \frac{(\lambda q)^{n_i} e^{-\lambda}}{{n_i}!} e^{\lambda (1-q)}  \nonumber\\
    &=\frac{(\lambda q)^{n_i} e^{-\lambda q}}{{n_i}!}  \nonumber\\
    &=P(n_i|\lambda q),
\end{align}
which is a Poisson distribution with mean $\lambda q = \lambda f(T,t_i)$.

\section{Bayesian multi-parameter estimation}\label{app:multi-parameter}

Our calculations assume that $\lambda$ in the likelihood model $p(n_i|T,t_i) = P[n_i|\lambda f(T,t_i)]$ is perfectly known. 
Yet, our value for $\lambda$ comes from calibration measurements, and its uncertainty has not been taken into account explicitly. 
While, as shown in Sec.~\ref{sec:res-a}, our results would remain unchanged even if $ \lambda $ were exactly known, here we summarise, for the sake of completeness, how our global--Bayesian technique can be generalised for multi-parameter estimation.

Firstly, we must think about the likelihood as a two-parameter model, i.e., $p(n_i|\lambda, T,t_i) = P[n_i|\lambda f(T,t_i)]$.
The corresponding posterior would then be constructed as
\begin{equation}
    p(\theta, \lambda |\boldsymbol{n},\boldsymbol{t}) \propto p(\theta, \lambda) \prod\nolimits_{i=1}^\mu p(n_i|\lambda,\theta, t_i),
    \label{eq:multi-posterior}
\end{equation}
in which we additionally allow for uncertainty in $\lambda$ through $p(\theta, \lambda)$.
$\lambda$ may be then treated as a nuisance parameter, which leads to the temperature estimator
\begin{equation*}
    \tilde{\vartheta}(\boldsymbol{n},\boldsymbol{t}) = \theta_u \exp\left[\int d\lambda d\theta \,p(\lambda,\theta|\boldsymbol{n},\boldsymbol{t}) \log{\left(\frac{\theta}{\theta_u}\right)}\right],
\end{equation*}
with mean logarithmic error 
\begin{equation*}
    \bar{\epsilon}_{\mathrm{mle}}(\boldsymbol{n},\boldsymbol{t}) = \int d\lambda d\theta\,p(\lambda, \theta|\boldsymbol{n},\boldsymbol{t}) \log^2{\left[\frac{\tilde{\vartheta}(\boldsymbol{n},\boldsymbol{t})}{\theta}\right]}.
\end{equation*}
If we instead wish to estimate $\lambda$, we can do so by marginalising over $\theta$ in Eq.~\eqref{eq:multi-posterior}.  

\bibliography{refs2022sep}

\end{document}